\documentclass[a4paper,USenglish]{article}
\usepackage[T1]{fontenc}
\usepackage{graphicx}
\usepackage[utf8]{inputenc}
\usepackage{amsmath, amsfonts, amssymb,mathtools,nicefrac, amsthm} 
\usepackage[shortlabels]{enumitem}
\usepackage[linesnumbered,ruled,vlined]{algorithm2e}
\SetKwRepeat{DoWhile}{do}{while}
\usepackage{caption}
\usepackage{subcaption}
\usepackage{url}
\usepackage[hidelinks]{hyperref}
\usepackage[capitalise]{cleveref}
\usepackage{placeins}
\usepackage{fontawesome}
\usepackage{tikz}
\usetikzlibrary{calc}
\usetikzlibrary{patterns}
\usetikzlibrary{decorations.pathreplacing}
\usetikzlibrary{fadings}
\usepackage{sectsty}
\allsectionsfont{\boldmath}

\theoremstyle{definition}
\newtheorem{definition}{Definition}
\theoremstyle{theorem}
\newtheorem{theorem}[definition]{Theorem}
\newtheorem{lemma}[definition]{Lemma}
\newtheorem{proposition}[definition]{Proposition}
\newtheorem{corollary}[definition]{Corollary}
\begin{document}
	\title{A $\nicefrac{4}{3}$-Approximation for the Maximum Leaf Spanning Arborescence Problem in DAGs}
	%
	\author{Meike Neuwohner}
	
	\date{}
	%
	%
	\maketitle              
	\begin{abstract}
	The Maximum Leaf Spanning Arborescence problem (MLSA) is defined as follows: Given a directed graph $G$ and a vertex $r\in V(G)$ from which every other vertex is reachable, find a spanning arborescence rooted at $r$ maximizing the number of leaves (vertices with out-degree zero). The MLSA has applications in broadcasting, where a message needs to be transferred from a source vertex to all other vertices along the arcs of an arborescence in a given network. In doing so, it is desirable to have as many vertices as possible that only need to receive, but not pass on messages since they are inherently cheaper to build.
	
	We study polynomial-time approximation algorithms for the MLSA.
	For general digraphs, the state-of-the-art is a $\min\{\sqrt{\mathrm{OPT}},92\}$-approxima-tion~\cite{DaligaultThomasse,Drescher2010AnAA}. In the (still APX-hard) special case where the input graph is acyclic, the best known approximation guarantee of $\frac{7}{5}$ is due to Fernandes and Lintzmayer~\cite{FernandesLintzmayer}: They prove that any $\alpha$-approximation for the \emph{hereditary $3$-set packing problem}, a special case of weighted $3$-set packing, yields a $\max\{\frac{4}{3},\alpha\}$-approximation for the MLSA in acyclic digraphs (dags), and provide a $\frac{7}{5}$-approximation for the hereditary $3$-set packing problem.
	
	In this paper, we obtain a $\frac{4}{3}$-approximation for the hereditary $3$-set packing problem, and, thus, also for the MLSA in dags. In doing so, we manage to leverage the full potential of the reduction provided by Fernandes and Lintzmayer. The algorithm that we study is a simple local search procedure considering swaps of size up to $10$. Its analysis relies on a two-stage charging argument.
	\end{abstract}
\newpage
\section{Introduction}
Given a (simple) directed graph $G=(V,E)$ and a root vertex $r\in V$, we call a subgraph $T$ of $G$ a \emph{spanning $r$-arborescence in $G$} if it satisfies the following conditions:
\begin{enumerate}[(i)]
	\item $T$ is a \emph{spanning subgraph} of $G$, that is, $V(T)=V$.
	\item $r$ does not have any entering arc in $T$ and each $v\in V\setminus \{r\}$ has exactly one entering arc in $T$.
	\item Each vertex in $V$ is reachable from $r$ via a directed path in $T$.
\end{enumerate}
We call a vertex $v$ a \emph{leaf} of $T$ if $v$ does not have any leaving arc in $T$.
\begin{figure}[t]
\begin{subfigure}{0.3\textwidth}
\centering
\scalebox{0.6}{	
\begin{tikzpicture}[mynode/.style={circle, minimum size = 2mm, inner sep = 0pt, draw, fill}, scale=0.5]
		\node[mynode, label=right:$r$] (R) at (13,5) {};
		\node[mynode] (H) at (16,0){};
		\node[mynode] (F) at (11,3){};
		\node[mynode] (A) at (6,4) {};
		\node[mynode] (B) at (6,0){};
		\node[mynode] (C) at (3,-3){};
		\node[mynode] (D) at (12,-5){};
		\node[mynode] (E) at (10,-1){};
		\node[mynode] (G) at (13,-1){};
		
		\draw[thick, ->] (R)--(F);
		\draw[thick, ->] (R)--(A);
		\draw[thick, ->] (F)--(E);
		\draw[thick, ->, bend left = 30] (F) to (G);
		\draw[thick, ->] (F)--(H);
		\draw[thick, ->] (A)--(B);
		\draw[thick, ->] (B)--(C);
		\draw[thick, ->] (B)--(D);
		
		\draw[thick, ->] (R)--(H);
		\draw[thick, ->] (H)--(G);
		\draw[thick, ->] (H)--(D);
		\draw[thick, ->, bend left = 30] (G) to (F);
		\draw[thick, ->] (D)--(E);
		\draw[thick, ->] (D)--(C);
		\draw[thick, ->] (E)--(B);
		\draw[thick, ->] (C)--(A);
		
		\draw[thick, ->] (F)--(A);
		\draw[thick, ->] (B)--(F);
		\draw[thick, ->] (A)--(E);
		\draw[thick, ->] (D)--(G);
		\draw[thick, ->] (C)--(G);
\end{tikzpicture}}
\end{subfigure}
\begin{subfigure}{0.3\textwidth}
\centering
\scalebox{0.6}{	
\begin{tikzpicture}[mynode/.style={circle, minimum size = 2mm, inner sep = 0pt, draw, fill}, scale=0.5]
		\node[mynode, label=right:$r$] (R) at (13,5) {};
		\node[mynode] (H) at (16,0){};
		\node (F) at (11,3){\color{green!70!black}\faLeaf};
		\node (A) at (6,4) {\color{green!70!black}\faLeaf};
		\node (B) at (6,0){\color{green!70!black}\faLeaf};
		\node[mynode] (C) at (3,-3){};
		\node[mynode] (D) at (12,-5){};
		\node[mynode] (E) at (10,-1){};
		\node[mynode] (G) at (13,-1){};
		
		\draw[thick, ->, black!50!white] (R)--(F);
		\draw[thick, ->, black!50!white] (R)--(A);
		\draw[thick, ->, black!50!white] (F)--(E);
		\draw[thick, ->, black!50!white, bend left = 30] (F) to (G);
		\draw[thick, ->, black!50!white] (F)--(H);
		\draw[thick, ->, black!50!white] (A)--(B);
		\draw[thick, ->, black!50!white] (B)--(C);
		\draw[thick, ->, black!50!white] (B)--(D);
		
		\draw[ultra thick, ->] (R)--(H);
		\draw[ultra thick, ->] (H)--(G);
		\draw[ultra thick, ->] (H)--(D);
		\draw[ultra thick, ->, bend left = 30] (G) to (F);
		\draw[ultra thick, ->] (D)--(E);
		\draw[ultra thick, ->] (D)--(C);
		\draw[ultra thick, ->] (E)--(B);
		\draw[ultra thick, ->] (C)--(A);
		
		\draw[thick, ->, black!50!white] (F)--(A);
		\draw[thick, ->, black!50!white] (B)--(F);
		\draw[thick, ->, black!50!white] (A)--(E);
		\draw[thick, ->, black!50!white] (D)--(G);
		\draw[thick, ->, black!50!white] (C)--(G);
\end{tikzpicture}}
\end{subfigure}
\begin{subfigure}{0.3\textwidth}
	\centering
	\scalebox{0.6}{
		\begin{tikzpicture}[mynode/.style={circle, minimum size = 2mm, inner sep = 0pt, draw, fill}, scale=0.5]
			\node[mynode, label=right:$r$] (R) at (13,5) {};
			\node (H) at (16,0){\color{green!70!black}\faLeaf};
			\node[mynode] (F) at (11,3){};
			\node[mynode] (A) at (6,4) {};
			\node[mynode] (B) at (6,0){};
			\node (C) at (3,-3){\color{green!70!black}\faLeaf};
			\node (D) at (12,-5){\color{green!70!black}\faLeaf};
			\node (E) at (10,-1){\color{green!70!black}\faLeaf};
			\node (G) at (13,-1){\color{green!70!black}\faLeaf};
			
			\draw[ultra thick, ->] (R)--(F);
			\draw[ultra thick, ->] (R)--(A);
			\draw[ultra thick, ->] (F)--(E);
			\draw[ultra thick, ->, bend left = 30] (F) to (G);
			\draw[ultra thick, ->] (F)--(H);
			\draw[ultra thick, ->] (A)--(B);
			\draw[ultra thick, ->] (B)--(C);
			\draw[ultra thick, ->] (B)--(D);
			
			\draw[thick, ->, black!50!white] (R)--(H);
			\draw[thick, ->, black!50!white] (H)--(G);
			\draw[thick, ->, black!50!white] (H)--(D);
			\draw[thick, ->, black!50!white, bend left = 30] (G) to (F);
			\draw[thick, ->, black!50!white] (D)--(E);
			\draw[thick, ->, black!50!white] (D)--(C);
			\draw[thick, ->, black!50!white] (E)--(B);
			\draw[thick, ->, black!50!white] (C)--(A);
			
			\draw[thick, ->, black!50!white] (F)--(A);
			\draw[thick, ->, black!50!white] (B)--(F);
			\draw[thick, ->, black!50!white] (A)--(E);
			\draw[thick, ->, black!50!white] (D)--(G);
			\draw[thick, ->, black!50!white] (C)--(G);
	\end{tikzpicture}}
\end{subfigure}
\caption{Illustration of the Maximum Leaf Spanning Arborescence problem. The leftmost picture shows a simple directed graph $G=(V,E)$, together with a vertex $r\in V$ from which every other vertex is reachable. The middle picture illustrates a spanning $r$-arborescence in $G$ with $3$ leaves (indicated by \textcolor{green!70!black}{\faLeaf}). The rightmost picture shows a  spanning $r$-arborescence in $G$ with $5$ leaves.}
\end{figure}
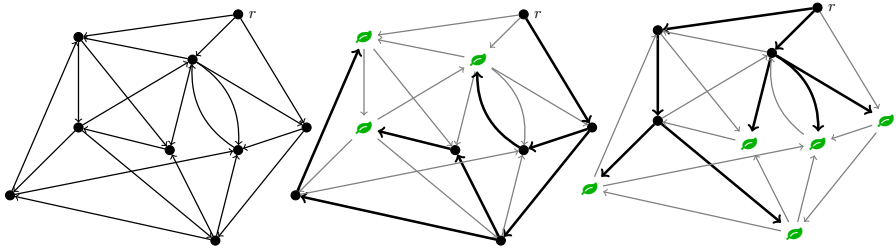
The Maximum Leaf Spanning Arborescence problem (MLSA) is defined as follows:
\begin{definition}[Maximum Leaf Spanning Arborescence problem]\
	\begin{description}
		\item[Input:] A directed graph $G$, $r\in V(G)$ such that every vertex of $G$ is reachable from $r$.
		\item[Task:] Find a spanning $r$-arborescence in $G$ with the maximum number of leaves possible.
	\end{description}
\end{definition}
It plays an important role in the context of broadcasting: Given a network consisting of a set of nodes containing one distinguished source and a set of available arcs, a message needs to be transferred from the source to all other nodes along a subset of the arcs, which forms (the arc set of) an arborescence rooted at the source. As internal nodes do not only need to be able to receive, but also to re-distribute messages, they are more expensive. Hence, it is desirable to have as few of them as possible, or equivalently, to maximize the number of leaves.

The special case of the MLSA where every arc may be used in both directions is called the \emph{Maximum Leaves Spanning Tree problem (MLST)}. In this setting, the complementary task of minimizing the number of non-leaves is equivalent to the \emph{Minimum Connected Dominating Set problem (MCDS)}. 
Both the MLST and the MCDS are NP-hard, even if the input graph is $4$-regular or planar with maximum degree at most $4$ (see~\cite{GareyJohnson}, problem ND2). 
The MLST has been shown to be APX-hard~\cite{GALBIATI199445}\footnote{Note that MaxSNP-hardness implies APX-hardness, see~\cite{OnSyntacticVersusComputationalViewsOfApproximability}.}, even when restricted to cubic graphs~\cite{BONSMA201214}. The state-of-the-art for the MLST is an approximation guarantee of $2$~\cite{SolisOba2015A2A}. 

While an optimum solution to the MLST gives rise to an optimum solution to the MCDS and vice versa, the MCDS turns out to be much harder to approximate:  Ruan et. al.~\cite{RUAN2004325} have obtained an $\ln \Delta + 2$-approximation, where $\Delta$ denotes the maximum degree in the graph. A reduction from Set Cover (with bounded set sizes) further shows that unless $\mathrm{P}=\mathrm{NP}$, the MCDS is hard to approximate within a factor of $\ln \Delta - \mathcal{O}(\ln\ln \Delta)$~\cite{Guha1998,Trevisan}. An analogous reduction further yields the same hardness result for the problem of computing a spanning arborescence with the minimum number of non-leaves in a rooted acyclic digraph of maximum out-degree $\Delta$.

In this paper, we study polynomial-time approximation algorithms for (a special case of) the MLSA. 
For general digraphs, the best that is known is a $\min\{\sqrt{\mathrm{OPT}},92\}$-approximation~\cite{DaligaultThomasse,Drescher2010AnAA}. Moreover, there is a line of research focusing on FPT-algorithms for the MLSA~\cite{SpanningDirectedTreesWithManyLeaves,KernelsForProblemsWithNoKernel,DaligaultThomasse}.

The special case where the graph $G$ is assumed to be a dag (directed acyclic graph) has been proven to be APX-hard by Schwartges, Spoerhase and Wolff~\cite{SchwartgesSpoerhaseWolff}. They further provided a $2$-approximation, which was then improved to $\frac{3}{2}$ by Fernandes and Lintzmayer~\cite{FERNANDES2022217}. Recently, the latter authors managed to enhance their approach to obtain a $\frac{7}{5}$-approximation~\cite{FernandesLintzmayer}, which has been unchallenged so far. In this paper, following the approach by Fernandes and Lintzmayer, we improve on these results and obtain a $\frac{4}{3}$-approximation for the MLSA in dags.

Fernandes and Lintzmayer~\cite{FernandesLintzmayer} tackle the MLSA in dags by reducing it, up to an approximation guarantee of $\frac{4}{3}$, to a special case of the  weighted $3$-set packing problem, which we call the \emph{hereditary $3$-set packing problem}. Fernandes and Lintzmayer~\cite{FernandesLintzmayer} prove it to be NP-hard via a reduction from $3$-Dimensional Matching~\cite{karp1972reducibility}.
\begin{definition}[weighted $k$-set packing problem]
\begin{description}\
	\item[Input:] A family $\mathcal{S}$ of sets, each of cardinality at most $k$, $w:\mathcal{S}\rightarrow\mathbb{R}_{\geq 0}$
	\item[Task:] Compute a disjoint sub-collection $A\subseteq \mathcal{S}$ maximizing the total weight $w(A)\coloneqq \sum_{s\in A} w(s)$. 
\end{description}	
\end{definition}
We call a set family $\mathcal{S}$ \emph{hereditary} if for every $s\in \mathcal{S}$, $\mathcal{S}$ contains all non-empty subsets of $s$.
\begin{definition}[hereditary $3$-set packing problem]
An instance of the \emph{hereditary $3$-set packing problem} is an instance $(\mathcal{S},w)$ of the weighted $3$-set packing problem, where $\mathcal{S}$ is a hereditary family and $w(s)=|s|-1$ for all $s\in \mathcal{S}$.\label{DefHereditary23SetPacking}
\end{definition}
As the weights can be deduced from the set sizes, we will omit them in the following and simply denote an instance of the hereditary $3$-set packing problem by $\mathcal{S}$ (instead of $(\mathcal{S},w)$).
\begin{theorem}[\cite{FernandesLintzmayer}]
	Let $\alpha\geq 1$ and assume that there is a polynomial-time $\alpha$-approximation algorithm for the hereditary $3$-set packing problem.
	Then there exists a polynomial-time $\max\{\alpha,\frac{4}{3}\}$-approximation for the MLSA in dags.\label{TheoFernandesLintzmayer}
\end{theorem}
For $k\leq 2$, the weighted $k$-set packing problem can be solved in polynomial time via a reduction to the Maximum Weight Matching problem~\cite{edmonds1965maximum}. In contrast, for $k\geq 3$, even the special case where $w\equiv 1$, the \emph{unweighted $k$-set packing problem}, is NP-hard because it generalizes the $3$-Dimensional Matching problem~\cite{karp1972reducibility}. The technique that has proven most successful in designing approximation algorithms for both the weighted and the unweighted $k$-set packing is \emph{local search}. Given a feasible solution $A$, we call a collection $X$ of pairwise disjoint sets a \emph{local improvement of $A$} if $w(X)>w(N(X,A))$, where
\[N(X,A)\coloneqq \{a\in A:\exists x\in X: a\cap x\neq \emptyset\}\] is the \emph{neighborhood of $X$ in $A$}. Note that $N(X,A)$ comprises precisely those sets that we need to remove from $A$ in order to be able to add the sets in $X$.  

The state-of-the-art is a $\min\{\frac{k+1-\tau_k}{2},0.4986\cdot (k+1)+0.0208\}$-approximation for the weighted $k$-set packing problem, where $\tau_k\geq 0.428$ for $k\geq 3$ and $\lim_{k\rightarrow\infty} \tau_k = \frac{2}{3}$~\cite{Neuwohner23,ThieryWard}. Note that the guarantee of $1.786$ for $k=3$ is worse than the guarantee of $\frac{7}{5}$ that Fernandes and Lintzmayer achieve for the hereditary $3$-set packing problem.

In order to obtain the approximation guarantee of $\frac{7}{5}$, Fernandes and Lintzmayer perform local search with respect to a modified weight function. In addition to certain improvements of constant size, they incorporate another, more involved class of local improvements that are related to alternating paths in a certain auxiliary graph. This makes the analysis more complicated because in addition to charging arguments similar to ours, more intricate considerations regarding the structure of the auxiliary graph are required.

In this paper, we study a local search algorithm that considers local improvements consisting of up to $10$ sets with respect to an objective that first maximizes the weight of the current solution, and second the number of sets of weight $2$ that are contained in it. We show that this algorithm yields a polynomial-time $\frac{4}{3}$-approximation for the hereditary $3$-set packing problem. In particular, this results in a polynomial-time $\frac{4}{3}$-approximation for the MLSA in dags. In doing so, we manage to tap the full potential of Theorem~\ref{TheoFernandesLintzmayer}. Moreover, this work serves as a starting point in identifying, understanding, and exploiting structural properties of set packing instances that arise naturally from other combinatorial problems. Studying these instance classes may ultimately turn reductions to set packing instances into a more powerful tool in the design of approximation algorithms.

The remainder of this paper is organized as follows:
In \cite{FernandesLintzmayer}, the reduction from the MLSA in dags to the hereditary $3$-set packing problem is done in an ad-hoc fashion, involving some pre-processing and several pages of analysis. As a result, the connection between the MLSA in dags and the hereditary $3$-set packing problem remains somewhat mysterious. 
Consequently, in \cref{sec:set_packing_in_disguise}, we point out that the MLSA in dags can be rephrased as a set packing problem in a simple and very natural way. In particular, this yields in an approximation-preserving reduction from the MLSA in dags to what we call the \emph{hereditary set packing problem}, a natural extension of the hereditary $3$-set packing problem to arbitrary set sizes. We further show that for every $k\geq 2$, an $\alpha$-approximation for the hereditary $k$-set packing problem, the restriction of the hereditary set packing problem to instances with sets of size at most $k$, implies a $\max\{\alpha,\frac{k+1}{k}\}$-approximation for the hereditary set packing problem, and thus, also the MLSA in dags.
 In doing so, we provide a clear picture of the connections between the MLSA in dags, the (general) hereditary set packing problem and the bounded size variants. Moreover, we obtain a significantly shortened and simplified, and, thus, arguably more intuitive proof of \cref{TheoFernandesLintzmayer}.

The lower bound of $\frac{k+1}{k}$ on the approximation guarantees that we can achieve for the MLSA in dags via a reduction to the hereditary $k$-set packing problem decreases with larger values of $k$. Hence, a natural question that arises is whether a better approximation ratio than $\frac{4}{3}$ can be achieved by reducing to the hereditary $k$-set packing problem with $k\geq 4$ instead. In \cref{sec:lower_bound}, we show, however, that this is not the case, at least if we restrict ourselves to the simple (but yet quite successful) algorithmic paradigm of local search with constant improvement size. More precisely, we show that an algorithm for the hereditary $k$-set packing problem that only considers local improvements of constant size cannot yield a better approximation ratio than $2-\frac{2}{k}$. Note that $k\mapsto\max\{\frac{k+1}{k},2-\frac{2}{k}\}$ has a unique minimum at $k=3$, where it attains a value of $\frac{4}{3}$. As such, the approximation guarantee of $\frac{4}{3}$ is optimal for the approach we consider.

 Finally, in \cref{Sec:Hereditary}, we present a simple local search based $\frac{4}{3}$-approximation for the hereditary $3$-set packing problem.

\section{A Set Packing Problem in Disguise \label{sec:set_packing_in_disguise}}
In this section, we point out that the MLSA in dags is, at its core, a set packing problem. In \cref{subsec:MLSA_to_hereditary}, we formally introduce the hereditary set packing problem and provide a simple approximation-preserving reduction from the MLSA in dags to it. In \cref{subsec:reduction_to_bounded_set_sizes}, we then show that up to an approximation guarantee of $\frac{k+1}{k}$, we can reduce further to a setting where all sets in our instance contain at most $k$ elements ($k\geq 1$). The special case $k=3$ yields a simple and self-contained proof of \cref{TheoFernandesLintzmayer}.
\subsection{Reducing the MLSA in DAGs to Hereditary Set Packing \label{subsec:MLSA_to_hereditary}}
The \emph{hereditary set packing problem} is defined as follows:
\begin{definition}[hereditary set packing problem]
\begin{description}
	\item[]
	\item[Input:] a hereditary set family $\mathcal{S}$
	\item[Task:] Compute a disjoint sub-collection $A\subseteq \mathcal{S}$ maximizing $w(A)=\sum_{s\in A} w(s)$, where $w(s)\coloneqq |s|-1$.
\end{description}
\end{definition}
In order to avoid an unnecessary, potentially exponential overhead in the encoding length, we will assume in the following that a hereditary set family $\mathcal{S}$ is implicitly given by only storing the inclusion-wise maximal sets in $\mathcal{S}$ explicitly.

Our main result for this section is given by the following theorem:
\begin{theorem}
Let $\alpha\geq 1$. If there is a polynomial-time $\alpha$-approximation algorithm for the hereditary set packing problem, then there is a polynomial-time $\alpha$-approximation algorithm for the MLSA in dags.	\label{theorem:MLSA_to_set_packing}
\end{theorem}
In order to phrase our reduction from the MLSA in dags to the hereditary set packing problem, we require the following definition:
\begin{definition}
	Let $G=(V,E)$ be a directed graph. For $v\in V$, we define $\Gamma_G^+(v)$ and $\Gamma_G^-(v)$ to be the set of out- and in-neighbors of $v$, respectively, that is,
	\[\Gamma_G^{+}(v)\coloneqq \{w\in V: (v,w)\in E\} \text{ and } \Gamma_G^{-}(v)\coloneqq \{w\in V: (w,v)\in E\}.\]
	If $G$ is clear from the context, we may omit the subscript $G$ and just write $\Gamma^+(v)$ and $\Gamma^-(v)$, respectively.
\end{definition}
The following proposition tells us that finding a spanning $r$-arborescence in $G$ can be interpreted as a set partitioning problem:
\begin{proposition}
	Let $(G=(V,E),r)$ be an instance of the MLSA in dags and let $T$ be a spanning subgraph of $G$. The following are equivalent:
	\begin{enumerate}[(a)]
		\item $T$ is a spanning $r$-arborescence in $G$.
		\item $\Gamma_T^-(r)=\emptyset$ and $|\Gamma_T^-(v)|=1$ for every $v\in V\setminus \{r\}$.
		\item The sets $(\Gamma_T^+(v))_{v\in V}$ form a partition of $V\setminus \{r\}$.
	\end{enumerate}	\label{prop:condition_arborescence}
\end{proposition}
\begin{proof}
	Clearly, (b) and (c) are equivalent. Moreover, by definition of a spanning $r$-arborescence, (a) implies (b). Hence, we are left with showing that any spanning subgraph $T$ of $G$ that complies with (b) constitutes a spanning $r$-arborescence in $G$. To this end, it remains to check that every vertex is reachable from $r$ via a directed path in $T$. But this follows from the fact that every vertex other than $r$ has an entering arc in $T$: As $G$ does not contain any directed cycle, we can simply follow the entering arcs backwards until we reach $r$.
\end{proof}
Moreover, it is easy to see that the number of leaves of a spanning $r$-arborescence $T$ can be expressed in terms of the sizes of the out-neighborhoods in $T$.
\begin{proposition}
	Let $T$ be an arborescence. Then the number of leaves of $T$ equals
	\[1+\sum_{v\in V(T): \Gamma^+_T(v)\neq \emptyset} (|\Gamma^+_T(v)|-1).\] \label{prop:number_of_leaves}
\end{proposition}
\begin{proof}
	The number of leaves of $T$ equals $|\{v\in V(T):\Gamma^+_T(v)=\emptyset\}|$. Using $|E(T)|=|V(T)|-1$, we calculate
	\begin{align*}0&=|V(T)|-|V(T)|\\&=1+|E(T)|-|V(T)|=1+\sum_{v\in V(T)} (|\Gamma_T^+(v)| - 1)\\
		&=1+\sum_{v\in V(T):\Gamma_T^+(v)\neq \emptyset} (|\Gamma_T^+(v)| - 1)-|\{v\in V(T):\Gamma^+_T(v)=\emptyset\}|.\end{align*}
	Rearranging yields the desired statement.
\end{proof}
\begin{figure}[t]
\centering
\begin{tikzpicture}[mynode/.style={circle, minimum size = 2mm, inner sep = 0pt, draw, fill}, xscale = 0.35, yscale=0.43]
	\node[mynode, label=left:$r$] (R) at (0,0){};
	\node[mynode] (A) at (-4,-3){};
	\node[mynode] (B) at  (4,-3){};
	
	\node[mynode] (C) at (-7,-6){};
	\node[mynode] (D) at  (-4,-6){};
	\node[mynode] (E) at  (-1,-6){};
	\node[mynode] (F) at  (2,-6){};
	\node[mynode] (G) at  (6,-6){};
	
	\node[mynode] (H) at (-9,-9){};
	\node[mynode] (I) at (-5,-9){};
	\node[mynode] (J) at (-1,-9){};
	\node[mynode] (K) at (2,-9){};
	
	\draw[thick,->,black!50!white] (A)--(B);
	\draw[thick,->,black!50!white] (D)--(B);
	\draw[thick,->,black!50!white] (D)--(K);
	\draw[thick,->,black!50!white] (I)--(H);
	\draw[thick,->,black!50!white] (J)--(I);
	\draw[thick,->,black!50!white] (D)--(H);
	\draw[thick,->,black!50!white] (K)--(J);
	\draw[thick,->,black!50!white] (F)--(J);
	\draw[thick,->,black!50!white, bend left=30] (K) to (I);
	\draw[thick,->,black!50!white] (F)--(G);
	\draw[thick,->,black!50!white] (G)--(K);
	\draw[thick,->,black!50!white, bend left=50] (R) to (G);
	
	\draw[ultra thick,->] (R)--(A);
	\draw[ultra thick,->] (R)--(B);
	
	\draw[ultra thick,->] (A)--(C);
	\draw[ultra thick,->] (A)--(D);
	\draw[ultra thick,->] (A)--(E);
	\draw[ultra thick,->] (B)--(F);
	\draw[ultra thick,->] (B)--(G);
	
	\draw[ultra thick,->] (C)--(H);
	\draw[ultra thick,->] (C)--(I);
	\draw[ultra thick,->] (E)--(J);
	\draw[ultra thick,->] (F)--(K);
	
	\draw[ultra thick,->, red!70!black] (R)--(A);
	\draw[ultra thick,->, red!70!black] (R)--(B);
	
	\draw[ultra thick,->, orange] (A)--(C);
	\draw[ultra thick,->, orange] (A)--(D);
	\draw[ultra thick,->, orange] (A)--(E);
	\draw[ultra thick,->, yellow!50!orange] (B)--(F);
	\draw[ultra thick,->, yellow!50!orange] (B)--(G);
	
	\draw[ultra thick,->, green!70!black] (C)--(H);
	\draw[ultra thick,->, green!70!black] (C)--(I);
	\draw[ultra thick,->, blue] (E)--(J);
	\draw[ultra thick,->, blue!50!red] (F)--(K);
	
	\node[mynode, draw=red!70!black, fill=red!70!black] at (0,0){};
	\draw[ultra thick, rounded corners, draw=red!70!black] (-4.5,-3.5) rectangle (4.5,-2.5);
	
	\node[mynode,draw=orange, fill = orange] at (-4,-3){};
	\draw[ultra thick, rounded corners, draw=orange] (-7.5,-6.5) rectangle (-0.5,-5.5);
	\node[mynode, draw=yellow!50!orange, fill=yellow!50!orange] (B) at  (4,-3){};
	\draw[ultra thick, rounded corners, draw=yellow!50!orange] (1.5,-6.5) rectangle (6.5,-5.5);
	
	\node[mynode, draw=green!70!black, fill=green!70!black] at (-7,-6){};
	\draw[ultra thick, rounded corners, draw=green!70!black] (-9.5,-9.5) rectangle (-4.5,-8.5);
	\node[mynode, draw=blue, fill=blue] (E) at  (-1,-6){};
	\draw[ultra thick, rounded corners, draw=blue] (-1.5,-9.5) rectangle (-0.5,-8.5);
	\node[mynode, draw=blue!50!red, fill=blue!50!red] (F) at  (2,-6){};
	\draw[ultra thick, rounded corners, draw=blue!50!red] (1.5,-9.5) rectangle (2.5,-8.5);
	
	\node at (0,-4){$\color{red!70!black}2-1$};
	\node at (-4,-7){$\color{orange}3-1$};
	\node at (4,-7){$\color{yellow!50!orange}2-1$};
	\node at (-7,-10){$\color{green!70!black}2-1$};
	\node at (-1,-10){$\color{blue}1-1$};
	\node at (4,-9){$\color{blue!50!red}1-1$};
\end{tikzpicture}
\caption{The figure illustrates a spanning $r$-arborescence $T$(bold arcs) in a directed graph $G=(V,E)$ (bold and gray arcs). The non-leaf vertices are marked in different colors and for each non-leaf, the leaving arcs are drawn in the same color. Moreover, colorful frames indicate the out-neighborhoods of the non-leafs. It can be seen that these form a partition of $V\setminus \{r\}$. The number of leaves of $T$ can be calculated by summing up the colorful numbers written below the out-neighborhoods (cf.\ \cref{prop:number_of_leaves}).}
\end{figure}
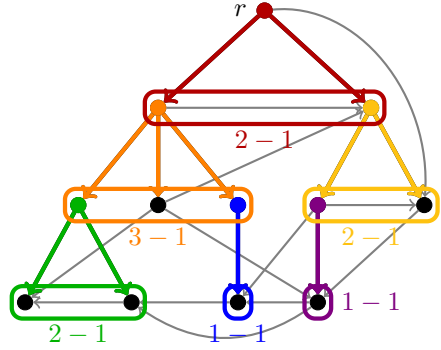
By \cref{prop:condition_arborescence} and \cref{prop:number_of_leaves}, finding a spanning $r$-arborescence with the maximum number of leaves is equivalent to partitioning $V\setminus \{r\}$ into a collection $\mathcal{S}$ of subsets of the out-neighborhoods of the vertices in $V$, maximizing the total weight $\sum_{s\in \mathcal{S}} (|s|-1)$. Given that adding additional elements to the sets cannot decrease the objective value, we may actually relax the condition that the sets in $\mathcal{S}$ \emph{partition} $V\setminus \{r\}$ to the weaker requirement that they are pairwise disjoint. This motivates the following definition:

\begin{definition}
Let $(G=(V,E),r)$ be an instance of the MLSA in dags. We define the \emph{hereditary set family associated with $G$} to be \[\mathcal{S}_G\coloneqq \{U\subseteq V:\exists v\in V:\emptyset\neq U\subseteq \Gamma^+_G(v)\}.\]
\end{definition}
Note that we can compute the inclusion-wise maximal sets in $\mathcal{S}_G$ in polynomial time $\mathcal{O}(|V|^3)$ by determining the inclusion-wise maximal ones among the sets $\{\Gamma^+_G(v):v\in V\}$.

In the following, we formally present the reduction from the MLSA in dags to the hereditary set packing problem. \cref{prop:arborescence_to_sets} shows that a spanning $r$-arborescence with $\ell$ leaves can be converted into a solution to $\mathcal{S}_G$ of objective value $\ell-1$. Conversely, \cref{lemma:sets_to_arborescence} tells us that given a solution to $\mathcal{S}_G$ of objective value $t$, we can, in polynomial-time, compute a spanning $r$-arborescence in $G$ with at least $t+1$ leaves.
\begin{proposition}
Let $(G=(V,E),r)$ be an instance of the MLSA in dags and let $T$ be a spanning $r$-arborescence in $G$ with $\ell$ leaves.

Define $A_T\coloneqq \{\Gamma^+_T(v):v\in V, \Gamma_T^+(v)\neq \emptyset\}$. Then $A_T$ is a feasible solution to $\mathcal{S}_G$ with objective value
$\sum_{s\in A_T} (|s|-1)=\ell-1$. \label{prop:arborescence_to_sets}
\end{proposition}
\begin{proof}
As in an arborescence, each vertex has at most $1$ entering arc, the sets in $A_T$ are pairwise disjoint. By \cref{prop:number_of_leaves}, we have
\[\sum_{s\in A_T} (|s|-1)=\sum_{v\in V: \Gamma^+_T(v)\neq \emptyset} (|\Gamma^+_T(v)|-1)=\ell-1.\]
\end{proof}
The following proposition is used to take care of the fact that the solution to $\mathcal{S}_G$ that we compute might not cover all vertices in $V\setminus\{r\}$.
\begin{proposition}
Let $G_1=(V,E_1)$ be a directed graph and let $G_2=(V,E_2)$ be a subgraph of $G_1$. Then
\[\sum_{v\in V:\Gamma^+_{G_1}(v)\neq \emptyset} (|\Gamma_{G_1}^+(v)|-1)\geq \sum_{v\in V:\Gamma^+_{G_2}(v)\neq \emptyset} (|\Gamma_{G_2}^+(v)|-1).\] \label{prop:monotonicity}
\end{proposition}
\begin{proof}
	By performing induction on $E_1\setminus E_2$, it suffices to consider the case where $E_1\setminus E_2$ consists of a single arc $e=(u,w)$. If $u$ has out-degree $0$ in $G_2$, then
	\[\sum_{v\in V:\Gamma^+_{G_1}(v)\neq \emptyset} (|\Gamma_{G_1}^+(v)|-1)=\sum_{v\in V:\Gamma^+_{G_2}(v)\neq \emptyset} (|\Gamma_{G_2}^+(v)|-1).\] Otherwise,
	\[\sum_{v\in V:\Gamma^+_{G_1}(v)\neq \emptyset} (|\Gamma_{G_1}^+(v)|-1)= 1+\sum_{v\in V:\Gamma^+_{G_2}(v)\neq \emptyset} (|\Gamma_{G_2}^+(v)|-1).\]
\end{proof}
\begin{lemma}
Let $(G=(V,E),r)$ be an instance of the MLSA in dags and let $A$ be a feasible solution to $\mathcal{S}_G$.
Then we can, in polynomial time, construct a spanning $r$-arborescence in $G$ with at least $1+\sum_{s\in A} (|s|-1)$ many leaves. \label{lemma:sets_to_arborescence}
\end{lemma}
\begin{proof}
For $s\in A$, pick $v_s$ such that $s\subseteq \Gamma_G^+(v)$. For $v\in V\setminus (\{r\}\cup \bigcup_{s\in A} s)$, pick an arbitrary entering arc $e_v\in\delta^-(v)$. Note that such an arc exists since every vertex is reachable from $r$ via a directed path in $G$.

Define a spanning subgraph $T$ of $G$ via $V(T)\coloneqq V$ and 
\[E(T)\coloneqq \{(v_s,w): w\in s\in A\}\cup \left\{e_v:v\in V\setminus \left(\{r\}\cup \bigcup_{s\in A} s\right)\right\}.\]	
By definition of $\mathcal{S}_G$, $T$ is a subgraph of $G$. As the sets in $A$ are pairwise disjoint, we have $|\Gamma^-_T(v)|=1$ for every $v\in V\setminus \{r\}$. Finally, as $G$ is acyclic and every vertex is reachable from $r$, $r$ does not have any in-neighbor in $G$. In particular, $\Gamma^-_{T}(r)=\emptyset$. By \cref{prop:condition_arborescence}, $T$ is a spanning $r$-arborescence in $G$. 

Denote by $T'$ the spanning subgraph of $T$ with arc set $E(T')\coloneqq \{(v_s,w): w\in s\in A\}$. By \cref{prop:number_of_leaves} and \cref{prop:monotonicity}, the number of leaves of $T$ can be lower bounded by
\begin{align*}&\phantom{=}1+\sum_{v\in V:\Gamma_{T'}^{+}(v)\neq \emptyset} (|\Gamma^+_{T'}(v)|-1)\\
&=1+\sum_{v\in V:\Gamma_{T'}^{+}(v)\neq \emptyset} |\Gamma^+_{T'}(v)|-|\{v\in V:\Gamma_{T'}^{+}(v)\neq \emptyset\}|\\
&=1+\sum_{s\in A} |s|-|\{v_s:s\in A\}| \geq 1+\sum_{s\in A} |s|-|A|=1+\sum_{s\in A} (|s|-1).\end{align*}
\end{proof}
Now, we are ready to prove \cref{theorem:MLSA_to_set_packing}.
\begin{proof}[Proof of \cref{theorem:MLSA_to_set_packing}]
Assuming a polynomial-time $\alpha$-approximation algorithm for the hereditary set packing problem, we obtain a polynomial-time $\alpha$-approximation for the MLSA in dags as follows:

For a given instance $(G,r)$, we first, in polynomial time, compute the representation of $\mathcal{S}_G$ by its inclusion-wise maximal sets. Next, we apply the $\alpha$-approximation algorithm for the hereditary set packing problem to obtain an $\alpha$-approximate solution $A$ to $\mathcal{S}_G$. Finally, we employ \cref{lemma:sets_to_arborescence} to construct a spanning $r$-arborescence $T$ in $G$ with at least $1+\sum_{s\in A}(|s|-1)$ many leaves.

In order to show that $T$ is an $\alpha$-approximate solution to the MLSA, denote the optimum value for $(G,r)$ by $\mathrm{OPT}$. Note that $\mathrm{OPT}\geq 1$. By \cref{prop:arborescence_to_sets}, there exists a feasible solution to $\mathcal{S}_G$ of objective value $\mathrm{OPT}-1$. As a consequence, we have 
\[\sum_{s\in A} (|s|-1) \geq \alpha^{-1}\cdot (\mathrm{OPT}-1).\]
This yields
\[1+\sum_{s\in A} (|s|-1)\geq \alpha^{-1}+\sum_{s\in A} (|s|-1)\geq \alpha^{-1}+\alpha^{-1}\cdot (\mathrm{OPT}-1)=\alpha^{-1}\cdot \mathrm{OPT}.\]
\end{proof}
\subsection{Reduction to Bounded Set Sizes \label{subsec:reduction_to_bounded_set_sizes}}
In this section, we show that for every $k\geq 1$, up to an approximation guarantee of $\frac{k+1}{k}$, we can reduce the hereditary set packing problem to the special case where all set sizes are bounded by $k$. The precise statement is given by \cref{theorem:hereditary_restrict_set_size}.
\begin{definition}[hereditary $k$-set packing problem]
The hereditary $k$-set packing problem is the restriction of the hereditary set packing problem to instances with sets of size at most $k$.
\end{definition}
Note that this definition coincides with \cref{DefHereditary23SetPacking} for $k=3$.
\begin{theorem}
Let $k\geq 1$. If there is a polynomial-time $\alpha$-approximation algorithm for the hereditary $k$-set packing problem, then there is a polynomial-time $\max\{\alpha,\frac{k+1}{k}\}$-approximation algorithm for the hereditary set packing problem. \label{theorem:hereditary_restrict_set_size}
\end{theorem}
Note that \cref{TheoFernandesLintzmayer} follows by combining \cref{theorem:MLSA_to_set_packing} and \cref{theorem:hereditary_restrict_set_size} for $k=3$.
\begin{proof}[Proof of \cref{theorem:hereditary_restrict_set_size}]
Assuming a polynomial-time $\alpha$-approximation algorithm for the hereditary $k$-set packing problem, we obtain a polynomial-time $\max\{\alpha,\frac{k+1}{k}\}$-approximation algorithm for the hereditary set packing problem as follows:

Given an instance $\mathcal{S}$ of the hereditary set packing problem, let \[\mathcal{S}_{\geq k+1}\coloneqq \{s\in \mathcal{S}: |s|\geq k+1\}.\]

As a first step, we compute a maximal solution $M\subseteq \mathcal{S}_{\geq k+1}$: To this end, we initialize $M=\emptyset$. We then traverse the inclusion-wise maximal sets in $\mathcal{S}$ in an arbitrary order. For each maximal set $s$, we check whether $|s\setminus \bigcup_{s'\in M}s'|\geq k+1$, and if yes, we add $s\setminus \bigcup_{s'\in M}s'$ to $M$.

We define $U\coloneqq \bigcup_{s\in M} s$. Let $\mathcal{S}'\coloneqq \{s\setminus U: s\in \mathcal{S}, s\setminus U\neq \emptyset\}$. By maximality of $M$, $\mathcal{S}'$ is an instance of the hereditary $k$-set packing problem. Moreover, the inclusion-wise maximal sets in $\mathcal{S}'$ are the inclusion-wise maximal ones among the sets $s\setminus U$, where $s\in\mathcal{S}$ is inclusion-wise maximal, and can, hence, be computed in polynomial time.

We apply the $\alpha$-approximation algorithm for the hereditary $k$-set packing problem to $\mathcal{S'}$ and obtain a solution $A'$. 

Finally, we output $A\coloneqq M\cup A'$.

By construction, the sets in $A$ are pairwise disjoint. Hence, it remains to prove that $A$ is a $\max\{\alpha,\frac{k+1}{k}\}$-approximate solution. To this end, let $B$ be an optimum solution for $\mathcal{S}$ and define $B'\coloneqq \{b\setminus U: b\in B, b\setminus U\neq \emptyset\}$. Then $B'$ is a feasible solution to $\mathcal{S'}$, which yields
\begin{equation}
\sum_{b\in B} |b\setminus U|-|B|\leq \sum_{b\in B} |b\setminus U|-|B'|=\sum_{b\in B'} (|b|-1)\leq \alpha\cdot \sum_{a\in A'} (|a|-1). \label{eq:bound_A_prime}
\end{equation} 
As the sets in $M$ are pairwise disjoint and of cardinality at least $k+1$, we obtain $\sum_{m\in M} |m|=|U|$ and $|M|\leq \frac{1}{k+1}\cdot |U|$. Using that the sets in $B$ are pairwise disjoint as well, we have
\begin{equation}
\sum_{b\in B} |b\cap U|\leq |U|=\frac{k+1}{k}\cdot (|U|-\frac{1}{k+1}\cdot |U|)\leq \frac{k+1}{k}\cdot \sum_{m\in M} (|m|-1). \label{eq:bound_M}
\end{equation}
Adding \eqref{eq:bound_A_prime} and \eqref{eq:bound_M} results in
\begin{align*}\sum_{b\in B} (|b|-1)&=\sum_{b\in B} |b\setminus U|-|B|+\sum_{b\in B} |b\cap U|\\
	&\leq \alpha\cdot \sum_{a\in A'} (|a|-1)+\frac{k+1}{k}\cdot \sum_{m\in M} (|m|-1)\\
	&\leq \max\left\{\alpha,\frac{k+1}{k}\right\}\cdot \sum_{a\in A} (|a|-1),\end{align*}
proving the desired approximation guarantee.
\end{proof}
\section{Lower Bound \label{sec:lower_bound}}
In this section, we show that we cannot obtain a better approximation guarantee than $2-\frac{2}{k}$ for the hereditary $k$-set packing problem via a local search algorithm that only considers local improvements of constant size. More precisely, we show that for every $k\geq 3$ and every $t\geq 1$, there exist (arbitrarily large) instances of the hereditary $k$-set packing problem that have a feasible solution $A$ that is locally optimum with respect to local improvements of size at most $t$, but whose weight is by a factor of at least $2-\frac{2}{k}$ smaller than the optimum. Note that a local search algorithm that iteratively searches for local improvements of size at most $t$ until no more exist might just pick $A$ set by set and then terminate.
\begin{theorem}
	Let $k\geq 3$ and $n,t\geq 1$. There exist 
	\begin{itemize}
		\item an instance $\mathcal{S}$ of the hereditary $k$-set packing problem with $|\mathcal{S}|\geq n$ and
		\item feasible solutions $A$ and $B$
	\end{itemize} 
	with the following properties:
	\begin{itemize}
		\item For every $X\subseteq \mathcal{S}\setminus A$ with $|X|\leq t$ and such that the sets in $X$ are pairwise disjoint, we have $w(X)<w(N(X,A))$. In particular, $A$ is locally optimum with respect to local improvements of size at most $t$.
		\item $w(B)=\left(2-\frac{2}{k}\right)\cdot w(A)$.
	\end{itemize}\label{theorem:lower_bound}
\end{theorem}
For the proof of \cref{theorem:lower_bound}, we first establish the following proposition, which is a direct consequence of a result by Erd{\H{o}}s and Sachs~\cite{ErdosSachs}.
\begin{proposition}
	Let $k\geq 3$ and $n,t\geq 1$. There is a simple $(2,k)$-regular bipartite graph $G$ with $|V(G)|\geq n$ and $\mathrm{girth}(G)\geq k\cdot t+1$, where $\mathrm{girth}(G)$ denotes the girth of $G$, i.e., the minimum length of a cycle in $G$. \label{prop:bipartite_graph_large_girth}
\end{proposition}
\begin{proof}
	Let $N\coloneqq \max\{n, (k-1)^{k\cdot t}\}$. By \cite{ErdosSachs}, there exists a $k$-regular graph $H$ on $|V(H)|\geq N$ vertices such that
	\[\mathrm{girth}(H)\geq \frac{\log(|V(H)|)}{\log(k-1)}-1\geq \frac{\log (N)}{\log(k-1)}-1\geq k\cdot t-1.\]
	Let $G$ be the bipartite vertex-edge-incidence graph of $H$, that is, 
	\[V(G)=V(H)\cup E(H) \text{ and } E(G)=\{\{v,e\}: v\in e\in E(H)\}.\]
	Then $G$ is a bipartite $(2,k)$-regular graph with $|V(G)|\geq |V(H)|\geq n$. As for every cycle $v_1,e_1,\dots,v_k,e_k$ in $G$ (where $v_1,\dots,v_k\in V(H)$ and $e_1,\dots,e_k\in E(H)$), $v_1,\dots,v_k$ is a cycle in $H$, we have \[\mathrm{girth}(G)\geq 2\cdot \mathrm{girth}(H)\geq 2\cdot k\cdot t-2\geq k\cdot t+1,\] where we used $k\geq 3$ and $t\geq 1$ for the last inequality.
\end{proof}
\begin{proof}[Proof of \cref{theorem:lower_bound}]
	Let $G=(V,E)$ be a simple $(2,k)$-regular bipartite graph with $|V|\geq n$ and $\mathrm{girth}(G)\geq k\cdot t+1$. Let $V_A$ and $V_B$ be the two bipartitions of $G$, where every vertex in $A$ has degree $2$, and every vertex in $B$ has degree $k$.
	
	Let $\mathcal{S}\coloneqq \{s\subseteq E: \exists v\in V: \emptyset\neq s\subseteq \delta(v)\}$ consist of the non-empty subsets of the sets of incident edges of vertices in $G$. As every vertex in $G$ has degree at most $k$, $\mathcal{S}$ is an instance of the hereditary $k$-set packing problem.
	
	Define $A\coloneqq \{\delta(v):v\in V_A\}$ and $B\coloneqq \{\delta(v):v\in V_B\}$. As $V_A$ and $V_B$ are independent sets in $G$, $A$ and $B$ both consist of pairwise disjoint sets. As every vertex in $V_A$ has degree $2$ and every vertex in $V_B$ has degree $k$, we have
	\[w(A)=\sum_{v\in V_A} (|\delta(v)|-1)=\frac{1}{2}\sum_{v\in V_A} |\delta(v)|=\frac{1}{2}\cdot |E|, \text{ and }\]
	\[w(B)=\sum_{v\in V_B} (|\delta(v)|-1)=\frac{k-1}{k}\sum_{v\in V_B} |\delta(v)|=\frac{k-1}{k}\cdot |E|.\]
	This yields $w(B)=\frac{2\cdot (k-1)}{k}\cdot w(A)=\left(2-\frac{2}{k}\right)\cdot w(A)$.
	
	It remains to show that $A$ is locally optimum. To this end, let $X\subseteq \mathcal{S}\setminus A$ such that the sets in $X$ are pairwise disjoint and $|X|\leq t$. We need to show that $w(X)<w(N(X,A))$.
	
	First of all, we may assume that $X$ does not contain any set $s\in\mathcal{S}$ with $|s|=1$ since $w(s)=0$ for such a set. In particular, as $X\subseteq \mathcal{S}\setminus A$ and $A=\{\delta(v):v\in V_A\}$ consists of sets of size $2$, we can infer that there is no $x\in X$ such that $x\subseteq \delta(v)$ for some $v\in V_A$.
	Consequently, for each $x\in X$, there is a (unique) $v_x\in V_B$ such that $x\subseteq \delta(v_x)$.
	
	Define $E_X\coloneqq \bigcup_{x\in X}x$ to be the collection of edges contained in the sets $x\in X$ and denote by $V_X\coloneqq \bigcup_{e\in E_X}e$ the set of endpoints of these edges. Then 
	\begin{equation}
		V_X\cap V_B=\{v_x:x\in X\} \text{ and } N(X,A)=\{\delta(v):v\in V_X\cap V_A\}.	\label{eq:VX}
	\end{equation}
	Using that all sets in $A$ have a size of $2$ and a weight of $1$, we can infer that
	\begin{equation}
		w(N(X,A))=|N(X,A)|=|V_X\cap V_A|. \label{eq:VX2}
	\end{equation} 
	
	As $|X|\leq t$, we know that $|E_X|\leq k\cdot |X|\leq k\cdot t$ and since the girth of $G$ is at least $k\cdot t+1$, $(V_X,E_X)$ is a forest. As such, we have 
	\begin{equation}
		|V_X|\geq |E_X|+1.\label{eq:forest}	
	\end{equation} Hence, we obtain
	\begin{align*}
		w(N(X,A))&\stackrel{\eqref{eq:VX2}}{=}|V_X\cap V_A|=|V_X|-|V_X\cap V_B|\stackrel{\eqref{eq:VX}}{\geq} |V_X|-|X|\\
		&\stackrel{\eqref{eq:forest}}{\geq} 1+|E_X|-|X|\stackrel{(*)}{=}1+\sum_{x\in X} (|x|-1)=1+w(X)>w(X),
	\end{align*}
	where the inequality marked $(*)$ follows from the fact that the sets in $X$ are pairwise disjoint.
\end{proof}
\section{\texorpdfstring{A $\nicefrac{4}{3}$-Approximation for the Hereditary $3$-Set Packing Problem}{A 4/3-Approximation for the Hereditary 3-Set Packing Problem}\label{Sec:Hereditary}}
In this section, we present a polynomial-time $\frac{4}{3}$-approximation for the hereditary $3$-set packing problem. For convenience, in the following, we will ignore the sets of size $1$ and weight $0$ contained in an instance $\mathcal{S}$ of the hereditary $3$-set packing problem because we can always remove them from any feasible solution without changing its weight.

 In order to phrase our algorithm, we formally introduce the notion of local improvement that we consider. It aims at maximizing first the weight of the solution we find, and second the number of sets of weight $2$ contained in it.
We first recap the notion of neighborhood from the introduction.
\begin{definition}[neighborhood]
Let $U$ and $W$ be two set families. We define the \emph{neighborhood} of $U$ in $W$ to be
\[N(U,W)\coloneqq \{w\in W:\exists u\in U: u\cap w\neq \emptyset\}.\]
Moreover, for a single set $u$, we write $N(u,W)\coloneqq N(\{u\},W)$.
\end{definition}
Now, we can define the notion of local improvement we would like to consider.
\begin{definition}[local improvement]
Let $\mathcal{S}$ be an instance of the hereditary $3$-set packing problem and let $A$ be a feasible solution. We call a disjoint set collection $X\subseteq \mathcal{S}$ a \emph{local improvement of $A$ of size $|X|$} if 
\begin{itemize}
	\item $w(X)>w(N(X,A))$ or
	\item $w(X)=w(N(X,A))$ and $X$ contains more sets of weight $2$ than $N(X,A)$.
\end{itemize}
\end{definition}
We analyze Algorithm~\ref{OverallAlgorithm}, which starts with the empty solution and iteratively searches for a local improvement of size at most $10$ (and performs the respective swap) until no more exists. We first observe that it runs in polynomial time.
\begin{proposition}\label{prop:poly_time}
Algorithm~\ref{OverallAlgorithm} can be implemented to run in polynomial time.
\end{proposition}
\begin{proof}
A single iteration can be performed in polynomial time via brute-force enumeration. Thus, it remains to bound the number of iterations. By our definition of a local improvement, $w(A)$ can never decrease throughout the algorithm. Initially, we have $w(A)=0$, and moreover, $w(A)\leq w(\mathcal{S})\leq 2\cdot|\mathcal{S}|$ holds throughout. As all weights are integral, we can infer that there are at most $2\cdot|\mathcal{S}|$ iterations in which $w(A)$ strictly increases. In between two consecutive such iterations, there can be at most $|\mathcal{S}|$ iterations in which $w(A)$ remains constant since the number of sets of weight $2$ in $A$ strictly increases in each such iteration. All in all, we can bound the total number of iterations by $\mathcal{O}(|\mathcal{S}|^2)$.
\end{proof}
\begin{algorithm}[t]
	\DontPrintSemicolon
	\KwIn{an instance $\mathcal{S}$ of the hereditary $3$-set packing problem}
	\KwOut{a disjoint sub-collection of $\mathcal{S}$}
$A\gets \emptyset$\;
	\While{$\exists$ local improvement $X$ of $A$ of size at most $10$}{
			$A\gets (A\setminus N(X,A))\cup X$
		}
	\textbf{return} $A$\;
	\caption{$\nicefrac{4}{3}$-approximation for hereditary $3$-set packing}\label{OverallAlgorithm}
\end{algorithm}
The remainder of this section is dedicated to the proof of Theorem~\ref{TheoMainHereditary}, which implies that Algorithm~\ref{OverallAlgorithm} constitutes a $\frac{4}{3}$-approximation for the hereditary $3$-set packing problem.
\begin{theorem}
	Let $\mathcal{S}$ be an instance of the hereditary $3$-set packing problem and let $A\subseteq \mathcal{S}$ be a feasible solution such that there is no local improvement of $A$ of size at most $10$. Let further $B\subseteq \mathcal{S}$ be an optimum solution. Then $w(B)\leq \frac{4}{3}\cdot w(A)$.\label{TheoMainHereditary}
\end{theorem}
 Let $\mathcal{S}$, $w$, $A$ and $B$ be as in the statement of the theorem. Our goal is to distribute the weights of the sets in $B$ among the sets in $A$ they intersect in such a way that no set in $A$ receives more than $\frac{4}{3}$ times its own weight. We remark that each set in $B$ must intersect at least one set in $A$ because otherwise, it would constitute a local improvement of size $1$. 

In order to present our weight distribution, we introduce the notion of the \emph{conflict graph}, which allows us to phrase our analysis using graph terminology. A similar construction is used in~\cite{FernandesLintzmayer}.
\begin{definition}[conflict graph]
The conflict graph $G$ is defined as follows: Its vertex set is the disjoint union of $A$ and $B$, i.e.\ $V(G)=A\dot{\cup}B$. Its edge set is obtained by adding, for each pair $(a,b)\in A\times B$, $|a\cap b|$ parallel edges connecting $a$ to $b$.
\end{definition} See \cref{FigConflictGraph} for an illustration. We remark that for $X\subseteq B$, $N(X,A)$ agrees with the (graph) neighborhood of $X$ in the bipartite graph $G$. Analogously, for $Y\subseteq A$, $N(Y,B)$ equals the neighborhood of $Y$ in $G$. In the following, we will simultaneously interpret sets from $A\dot{\cup} B$ as the corresponding vertices in $G$ and talk about their degree, their incident edges and their neighbors. We make the following observation.
\begin{proposition}
	Let $v\in V(G)$ correspond to the set $s\in A\cup B$. Then $v$ has at most $|s|$ incident edges in $G$.\label{prop:degree_in_conflict_graph}
\end{proposition}
\begin{proof}
	As $A$ and $B$ both consist of pairwise disjoint sets, each element of $s$ can induce at most one incident edge of $v$.
\end{proof}
\begin{figure}[t]
	\begin{subfigure}[t]{0.45\textwidth}
		\centering
		\begin{tikzpicture}[scale = 1.2,elem/.style = {circle, draw = black, fill, inner sep = 0.5mm, minimum size = 2mm}]
		\node[elem] (A) at (0,1) {};
		\node[elem] (B) at (1,1) {};
		\node[elem] (C) at (2,1) {};
		\node[elem] (D) at (0.5,0) {};
		\node[elem] (E) at (2,0) {};
		\draw[blue, thick, rounded corners] (-0.5,0.5) rectangle (2.5,1.5);
		\draw[red!70!black, ultra thick, rounded corners, dashed] (1.6,-0.4) rectangle (2.4,1.4);
		\draw[rounded corners = 5mm, red!70!black, ultra thick, dashed] (-0.5,1.4)--(1.5,1.4)--(0.5,-0.5)--cycle;
		\node at (3,1) {\color{blue}$A$};
		\node at (3,0) {\color{red!70!black}$B$};
		\end{tikzpicture}
		\caption{The figure displays two collections $A$ (blue, solid) and $B$ (red, dashed) consisting of pairwise disjoint sets of cardinality $2$ or $3$. Black dots represent set elements.}
	\end{subfigure}
	\quad
	\begin{subfigure}[t]{0.45\textwidth}
		\centering
		\begin{tikzpicture}[scale = 1.2,anode/.style = {circle, draw = blue, thick, fill = none, inner sep = 0mm, minimum size = 5mm},bnode/.style = {circle, draw = red!70!black, thick, fill = none, inner sep = 0mm, minimum size = 5mm}]
		\node[anode] (A2) at (1,1){};
		\node[bnode] (B2) at (0.5,0){};
		\node[bnode] (B3) at (2,0){};
		\node at (3,1) {\color{blue}$A$};
		\node at (3,0) {\color{red!70!black}$B$};
		\draw[thick] (A2) to[bend left = 30] (B2);
		\draw[thick] (A2) to[bend right = 30] (B2);
		\draw[thick] (A2)--(B3);
		\end{tikzpicture}
		\caption{The figure shows the conflict graph of $A\dot{\cup} B$. Vertices from $A$ are drawn in blue at the top, vertices from $B$ are drawn in red at the bottom.}
	\end{subfigure}
	\caption{Construction of the conflict graph. \label{FigConflictGraph}}
\end{figure}
\subsection{Step 1 of the Weight Distribution}
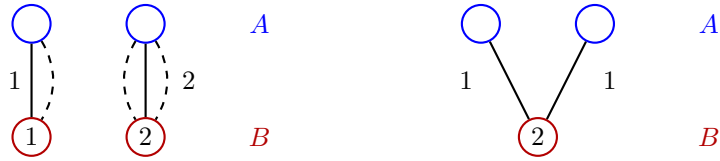
\begin{figure}[t]
	\begin{subfigure}[t]{0.45\textwidth}
		\centering
		\begin{tikzpicture}[scale = 1.5,anode/.style = {circle, draw = blue, thick, fill = none, inner sep = 0mm, minimum size = 5mm},bnode/.style = {circle, draw = red!70!black, thick, fill = none, inner sep = 0mm, minimum size = 5mm}]
		\node[anode] (A1) at (0,1){};
		\node[anode] (A2) at (1,1){};
		\node[bnode] (B1) at (0,0){$1$};
		\node[bnode] (B2) at (1,0){$2$};
		\node at (2,1) {\color{blue}$A$};
		\node at (2,0) {\color{red!70!black}$B$};
		\draw[thick] (A1) to node[midway, left] {$1$}(B1);
		\draw[thick, dashed] (A1) to[bend left = 30] (B1);
		\draw[thick, dashed] (A2) to[bend left = 30] (B2);
		\draw[thick, dashed] (A2) to[bend right = 30] (B2);
		\draw[thick] (A2) to node[midway, right=10pt] {$2$}(B2);
		\end{tikzpicture}
		\caption{Every set in $B_1$ sends its whole weight to its unique neighbor in $A$ (to which it may be connected via multiple edges).}
	\end{subfigure}
	\quad
	\begin{subfigure}[t]{0.45\textwidth}
		\centering
		\begin{tikzpicture}[scale = 1.5,anode/.style = {circle, draw = blue, thick, fill = none, inner sep = 0mm, minimum size = 5mm},bnode/.style = {circle, draw = red!70!black, thick, fill = none, inner sep = 0mm, minimum size = 5mm}]
		\node[anode] (A1) at (0,1){};
		\node[anode] (A2) at (1,1){};
		\node[bnode] (B1) at (0.5,0){$2$};
		\node at (2,1) {\color{blue}$A$};
		\node at (2,0) {\color{red!70!black}$B$};
		\draw[thick] (A1) to node[midway, left=10pt] {$1$}(B1);
		\draw[thick] (A2) to node[midway, right=10pt] {$1$}(B1);
		\end{tikzpicture}
		\caption{Every set in $B_2$ sends one unit of weight to each of its neighbors in $A$.}
	\end{subfigure}
	\caption{The first step of the weight distribution.\label{FigStep1}}
\end{figure}

Our weight distribution proceeds in two steps. The first step works as follows:

\begin{definition}[Step 1 of the weight distribution]
Let $B_1$ consist of all sets $v\in B$ with \emph{exactly one neighbor} in $A$. Each $v\in B_1$ sends its full weight to its unique neighbor in $A$.

Let further $B_2$ consist of those $v\in B$ with $w(v)=2$ and \emph{exactly two incident edges}, with the additional property that they connect to \emph{two distinct sets} from $A$. Each $v\in B_2$ sends half of its weight (i.e., $1$) along each of its edges.
\end{definition} See \cref{FigStep1} for an illustration. Observe that in the first stage, $u\in A$ receives weight precisely from the sets in $N(u,B_1\cup B_2)$.

We first prove Lemma~\ref{LemFirstStep}, which tells us that we can represent the total amount of weight a collection $U\subseteq A$ receives in the first step as the weight of a disjoint set collection $X$ with $N(X,A)\subseteq U$. The construction of $X$ will allow us to combine $X$ with sub-collections of $B\setminus(B_1\cup B_2)$ to obtain local improvements.
\begin{lemma}
	Let $U\subseteq A$. There is $X\subseteq \mathcal{S}$ with the following properties:
	\begin{enumerate}[label=(\thelemma.\arabic*)]
		\item \label{FirstStepProp1} $N(X,A)\subseteq U$.
		\item \label{FirstStepProp3} $w(X)$ equals the total amount of weight that $U$ receives in the first step.
		\item \label{FirstStepProp2} There is a bijection $N(U,B_1\cup B_2)\leftrightarrow X$ mapping $v\in B_1\cup B_2$ to itself or to one of its two-element subsets.
	\end{enumerate} \label{LemFirstStep}
\end{lemma}
\begin{figure}[h]
	\begin{subfigure}[t]{\textwidth}
		\centering
		\begin{tikzpicture}[scale = 1.2,elem/.style = {circle, draw = black, fill, inner sep = 0.5mm, minimum size = 2mm}]
		\node[elem] (A) at (0,1) {};
		\node[elem] (B) at (1,1) {};
		\node[elem] (C) at (2,1) {};
		\node[elem] at (3,1) {};
		\node[elem] at (4,1) {};
		\node[elem] (I) at (-1,1) {};
		\node[elem] (J) at (-2,1) {};
		\node[elem] (H) at (-3,1) {};
		\node[elem] (K) at (-3,0) {};
		\node[elem] (D) at (-0.5,0) {};
		\node[elem] (E) at (2.5,0) {};
		\draw[blue, thick, rounded corners, fill = blue, fill opacity = 0.1] (-0.4,0.5) rectangle (2.4,1.5);
		\draw[blue, thick, rounded corners, fill = blue, fill opacity = 0.1] (-3.5,0.5) rectangle (-0.5,1.5);
		\draw[blue, thick, rounded corners] (2.5,0.5) rectangle (4.5,1.5);
		\draw[rounded corners = 5mm, red!70!black, ultra thick, dashed] (-1.5,1.44)--(0.5,1.44)--(-0.5,-0.5)--cycle;
		\draw[rounded corners = 5mm, red!70!black, ultra thick, dashed] (1.5,1.44)--(3.5,1.45)--(2.5,-0.5)--cycle;
		\draw[rounded corners, red!70!black, ultra thick, dashed] (-3.4,-0.4) rectangle (-2.6,1.4);
		\node at (-4,1) {\color{blue}$A$};
		\node at (-4,0) {\color{red!70!black}$B$};
		\end{tikzpicture}
		\caption{The left red set is contained in $B_1$ and sends its whole weight to the unique set from $A$ it intersects. The two triangular red sets are contained in $B_2$. The left one only intersects sets in $A$ that are contained in $U$, whereas the right one also intersects a set in $A\setminus U$. \label{SubFigLemSetConfiguration} }
	\end{subfigure}

\vspace{5pt}

	\begin{subfigure}[t]{\textwidth}
		\centering
		\begin{tikzpicture}[scale = 1.2,elem/.style = {circle, draw = black, fill, inner sep = 0.5mm, minimum size = 2mm}]
		\node[elem] (A) at (0,1) {};
		\node[elem] (B) at (1,1) {};
		\node[elem] (C) at (2,1) {};
		\node[elem] at (3,1) {};
		\node[elem] at (4,1) {};
		\node[elem] (I) at (-1,1) {};
		\node[elem] (J) at (-2,1) {};
		\node[elem] (H) at (-3,1) {};
		\node[elem] (K) at (-3,0) {};
		\node[elem] (D) at (-0.5,0) {};
		\node[elem] (E) at (2.5,0) {};
		\draw[blue, thick, rounded corners, fill = blue, fill opacity = 0.1] (-0.4,0.5) rectangle (2.4,1.5);
		\draw[blue, thick, rounded corners, fill = blue, fill opacity = 0.1] (-3.5,0.5) rectangle (-0.5,1.5);
		\draw[blue, thick, rounded corners] (2.5,0.5) rectangle (4.5,1.5);
		\draw[rounded corners = 5mm, red!70!black, ultra thick, dashed] (-1.5,1.44)--(0.5,1.44)--(-0.5,-0.5)--cycle;
		\draw[rounded corners, red!70!black, ultra thick, dashed] (-3.4,-0.4) rectangle (-2.6,1.4);
		\node at (-4,1) {\color{blue}$A$};
		\node at (-4,0) {\color{red!70!black}$B$};
		\draw[rounded corners, red!70!black, ultra thick, dashed] (1.9-0.2,1.2-0.1)--(1.9+0.2,1.2+0.1)--(2.6+0.2,-0.2+0.1)--(2.6-0.2,-0.2-0.1)--cycle;
		\end{tikzpicture}
		\caption{The set collection $X$ (red, dashed) we construct in the proof of Lemma~\ref{LemFirstStep} contains the left and the middle red set because they send all of their weight to $U$. For the right triangular set, we remove the element in which it intersects a set from $A\setminus U$. Then, we add the resulting set of cardinality $2$ to $X$.\label{SubFigLemX}}
	\end{subfigure}
	\caption{Illustration of the construction in the proof of Lemma~\ref{LemFirstStep}. \cref{SubFigLemSetConfiguration} shows a collection $U\subseteq A$ of sets (blue, filled, solid), the collection $N(U,B_1\cup B_2)$ (red, dashed) of sets the sets in $U$ receive weight from in the first step, and further sets from $A$ (blue, not filled, solid) the sets in $N(U,B_1\cup B_2)$ send weight to. \cref{SubFigLemX} illustrates the construction of the set collection $X$.}\label{FigProofLemFirstStep}
\end{figure}
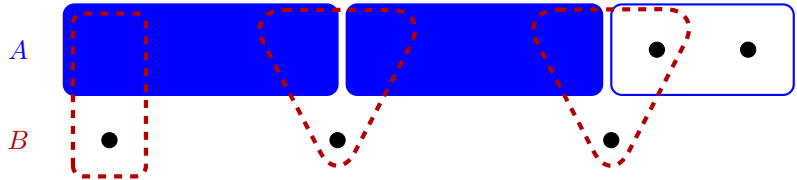
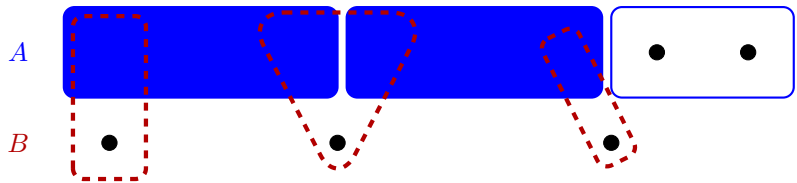
\begin{proof}
	We obtain $X$ as follows: We start with $X=\emptyset$ and first add those sets in $N(U,B_1\cup B_2)$ to $X$ that send all of their weight to $U$ (i.e., whose neighborhood in $A$ is contained in $U$). This includes all sets in $N(U,B_1)$. Second, for each set $v\in B_2$ that has one incident edge to $u\in U$ and one incident edge to $r\in A\setminus U$, we add its two-element subset $v\setminus r$ to $X$. By construction, \ref{FirstStepProp1}-\ref{FirstStepProp2} hold. See \cref{FigProofLemFirstStep} for an illustration.
\end{proof}
\begin{corollary}
	No set in $A$ receives more than its own weight in the first step.\label{CorNoMoreThanOwnWeight}
\end{corollary}
\begin{proof}
	Assume towards a contradiction that $u\in A$ receives more than $w(u)$ in the first step.  Apply Lemma~\ref{LemFirstStep} with $U=\{u\}$ to obtain a collection $X\subseteq \mathcal{S}$ subject to \ref{FirstStepProp1}-\ref{FirstStepProp2}. Then $w(X)>w(u)=w(N(X,A))$ and \ref{FirstStepProp2} and \cref{prop:degree_in_conflict_graph} imply that $X$ is a disjoint set family with $|X|\leq 3$. Hence, $X$ constitutes a local improvement of size at most $3<10$. This contradicts our assumption that there is no local improvement of $A$ of size at most $10$.
\end{proof}
\subsection{Removing ``Covered'' Sets}
\begin{definition}
	Let $C$ consist of those sets from $A$ that receive exactly their own weights in the first step.
\end{definition}
The intuitive idea behind our analysis is that the sets in $C$ are ``covered'' by the sets sending weight to them in the sense of \cref{LemFirstStep}. Hence, we can ``remove'' the sets in $C$ from our current solution $A$ and the sets in $B_1\cup B_2$ from our optimum solution $B$. If we can find a local improvement in the remaining instance, we will use Lemma~\ref{LemFirstStep} to transform it into a local improvement in the original instance, leading to a contradiction. See \cref{cor:single_vertex_improvement_without_C} for an example of how to apply this reasoning. But under the assumption that no local improvement in the remaining instance exists, we can design the second step of the weight distribution in such a way that overall, no set in $A$ receives more than $\frac{4}{3}$ times its own weight.
\subsection{Step 2 of the Weight Distribution}
In order to define the second step of the weight distribution, we make the following observations:
\begin{lemma}
	There is no $v\in B\setminus (B_1\cup B_2)$ with $w(N(v,A\setminus C))<w(v)$.\label{cor:single_vertex_improvement_without_C}
\end{lemma}
\begin{proof}
	Assume towards a contradiction that there is $v\in B\setminus (B_1\cup B_2)$ with $w(N(v,A\setminus C))<w(v)$.
Apply \cref{LemFirstStep} to $U\coloneqq N(v,C)$ to obtain $X$ subject to \ref{FirstStepProp1}-\ref{FirstStepProp2}. By \ref{FirstStepProp2}, $X\dot{\cup} \{v\}$ consists of pairwise disjoint sets. \cref{prop:degree_in_conflict_graph} further yields $|N(v,C)|\leq |v|\leq 3$, and, thus, $|X|=|N(N(v,C),B_1\cup B_2)|\leq 9$ by \ref{FirstStepProp2}. Finally, $w(X)=w(N(v,C))$ by \ref{FirstStepProp3} and since sets from $C$ receive exactly their own weights in the first step. Hence, \ref{FirstStepProp1} yields
\[w(X\cup \{v\})=w(X)+w(v) > w(N(v,C))+w(N(v,A\setminus C)) = w(N(X\cup \{v\},A)).\] So $X\cup \{v\}$ is a local improvement of $A$ of size at most $10$, a contradiction.
\end{proof}
\begin{proposition}
	Let $v\in B\setminus (B_1\cup B_2)$. Then
	\begin{enumerate}[(i)]
		\item $v$ has at least one neighbor in $A\setminus C$.
		\item If $w(v)=1$, then $v$ has exactly two neighbors in $A$.
		\item If $w(v)=2$, then $v$ has three incident edges.
	\end{enumerate}
\end{proposition}
\begin{proof}
$(i)$ follows from \cref{cor:single_vertex_improvement_without_C}. For $(ii)$ and $(iii)$, we remind ourselves that each $v\in B$ has at most $|v|$ neighbors/incident edges, but at least $1$ neighbor in $A$ by \cref{prop:degree_in_conflict_graph} and since $\{v\}$ would constitute a local improvement otherwise. $(ii)$ holds since $v\in B_1$ otherwise. For $(iii)$, we observe that in case $v$ has at most $2$ incident edges, then either $v$ has only one neighbor in $A$, or two distinct neighbors to which it is connected by a single edge each. In either case, we have $v\in B_1\cup B_2$.
\end{proof}

\begin{figure}
	\begin{subfigure}{0.19\textwidth}
		\centering
		\begin{tikzpicture}[scale = 1.4,anode/.style = {circle, draw = blue, thick, fill = none, inner sep = 0mm, minimum size = 5mm},bnode/.style = {circle, draw = red!70!black, thick, fill = none, inner sep = 0mm, minimum size = 5mm}]
		\node[anode, dashed] (C) at (0,1){};
		\node[anode] (A) at (1,1){};
		\node[bnode] (B) at (0.5,0){$1$};
		\draw[thick] (B) to node[midway, left]{$\frac{1}{3}$} (C);
		\draw[thick] (B) to node[midway, right]{$\frac{2}{3}$} (A);
		\end{tikzpicture}
		\caption*{\ref{SituationA}}
	\end{subfigure}
	\begin{subfigure}{0.19\textwidth}
		\centering
		\begin{tikzpicture}[scale = 1.4,anode/.style = {circle, draw = blue, thick, fill = none, inner sep = 0mm, minimum size = 5mm},bnode/.style = {circle, draw = red!70!black, thick, fill = none, inner sep = 0mm, minimum size = 5mm}]
		\node[anode] (C) at (0,1){};
		\node[anode] (A) at (1,1){};
		\node[bnode] (B) at (0.5,0){$1$};
		\draw[thick] (B) to node[midway, left]{$\frac{1}{2}$} (C);
		\draw[thick] (B) to node[midway, right]{$\frac{1}{2}$} (A);
		\end{tikzpicture}
		\caption*{\ref{SituationB}}
	\end{subfigure}
	\begin{subfigure}{0.3\textwidth}
		\centering
		\begin{tikzpicture}[scale = 1.4,anode/.style = {circle, draw = blue, thick, fill = none, inner sep = 0mm, minimum size = 5mm},bnode/.style = {circle, draw = red!70!black, thick, fill = none, inner sep = 0mm, minimum size = 5mm}]
		\node[anode, dashed] (C1) at (0,1){};
		\node[anode, dashed] (C2) at (-1,1){};
		\node[anode] (A) at (1,1){$2$};
		\node[bnode] (B) at (0.5,0){$2$};
		\draw[thick] (B) to node[midway, left]{$\frac{1}{3}$} (C1);
		\draw[thick] (B) to node[midway, left, below = 0.5pt]{$\frac{1}{3}$} (C2);
		\draw[thick] (B) to node[midway, right]{$\frac{4}{3}$} (A);
		\end{tikzpicture}
		\caption*{\ref{SituationC}}
	\end{subfigure}
	\begin{subfigure}{0.3\textwidth}
		\centering
		\begin{tikzpicture}[scale = 1.4,anode/.style = {circle, draw = blue, thick, fill = none, inner sep = 0mm, minimum size = 5mm},bnode/.style = {circle, draw = red!70!black, thick, fill = none, inner sep = 0mm, minimum size = 5mm}]
		\node[anode] (A1) at (0,1){$1$};
		\node[anode, dashed] (C) at (-1,1){};
		\node[anode] (A2) at (1,1){$1$};
		\node[bnode] (B) at (0.5,0){$2$};
		\draw[thick] (B) to node[midway, left]{$\frac{2}{3}$} (A1);
		\draw[thick] (B) to node[midway, left, below = 0.5pt]{$\frac{2}{3}$} (C);
		\draw[thick] (B) to node[midway, right]{$\frac{2}{3}$} (A2);
		\end{tikzpicture}
		\caption*{\ref{SituationD}}
	\end{subfigure}
	\begin{subfigure}{0.3\textwidth}
		\centering
		\begin{tikzpicture}[scale = 1.4,anode/.style = {circle, draw = blue, thick, fill = none, inner sep = 0mm, minimum size = 5mm},bnode/.style = {circle, draw = red!70!black, thick, fill = none, inner sep = 0mm, minimum size = 5mm}]
		\node[anode] (A1) at (0,1){$1$};
		\node[anode, dashed] (C) at (-1,1){};
		\node[anode] (A2) at (1,1){$2$};
		\node[bnode] (B) at (0.5,0){$2$};
		\draw[thick] (B) to node[midway, left]{$\frac{2}{3}$} (A1);
		\draw[thick] (B) to node[midway, left, below = 0.5pt]{$\frac{1}{3}$} (C);
		\draw[thick] (B) to node[midway, right]{$1$} (A2);
		\end{tikzpicture}
		\caption*{\ref{SituationD}}
	\end{subfigure}
	\begin{subfigure}{0.3\textwidth}
		\centering
		\begin{tikzpicture}[scale = 1.4,anode/.style = {circle, draw = blue, thick, fill = none, inner sep = 0mm, minimum size = 5mm},bnode/.style = {circle, draw = red!70!black, thick, fill = none, inner sep = 0mm, minimum size = 5mm}]
		\node[anode] (A1) at (0,1){$2$};
		\node[anode, dashed] (C) at (-1,1){};
		\node[anode] (A2) at (1,1){$2$};
		\node[bnode] (B) at (0.5,0){$2$};
		\draw[thick] (B) to node[midway, left]{$1$} (A1);
		\draw[thick] (B) to node[midway, left, below = 0.5pt]{$0$} (C);
		\draw[thick] (B) to node[midway, right]{$1$} (A2);
		\end{tikzpicture}
		\caption*{\ref{SituationD}}
	\end{subfigure}
	\begin{subfigure}{0.3\textwidth}
		\centering
		\begin{tikzpicture}[scale = 1.4,anode/.style = {circle, draw = blue, thick, fill = none, inner sep = 0mm, minimum size = 5mm},bnode/.style = {circle, draw = red!70!black, thick, fill = none, inner sep = 0mm, minimum size = 5mm}]
		\node[anode] (A1) at (0,1){};
		\node[anode] (A3) at (-1,1){};
		\node[anode] (A2) at (1,1){};
		\node[bnode] (B) at (0,0){$2$};
		\draw[thick] (B) to node[midway, left]{$\frac{2}{3}$} (A1);
		\draw[thick] (B) to node[midway, left, below = 0.5pt]{$\frac{2}{3}$} (A3);
		\draw[thick] (B) to node[midway, right, below = 0.5pt]{$\frac{2}{3}$} (A2);
		\end{tikzpicture}
		\caption*{\ref{SituationE}}
	\end{subfigure}
	\caption{Illustration of the second step of the weight distribution. Blue circles in the top row indicate sets from $A$, if they are dashed, the corresponding set is contained in $C$. Red circles in the bottom row indicate sets from $B\setminus (B_1\cup B_2)$. The number within a circle indicates the weight of the corresponding set in case it is relevant. Even though drawn as individual circles, the endpoints in $A$ of the incident edges of a set $v\in B\setminus(B_1\cup B_2)$ need not be distinct. For example, in \ref{SituationE}, two of the sets represented by the blue circles may agree, in which case the corresponding set receives $\frac{4}{3}$.\label{FigSecondStep}}
\end{figure}
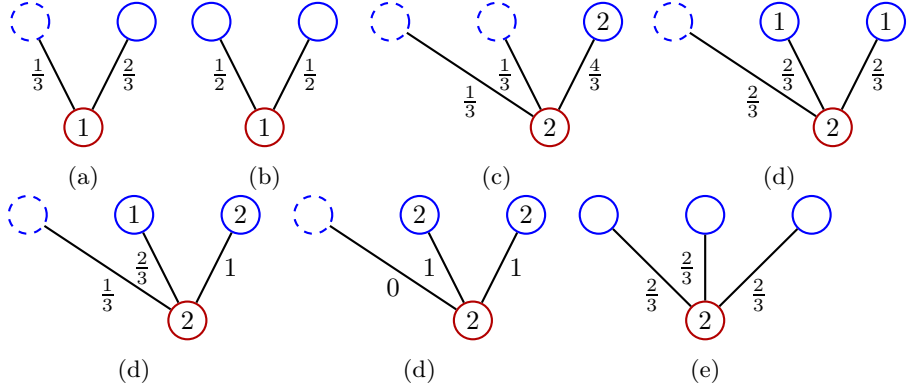

\begin{definition}[Step 2 of the weight distribution]
Let $v\in B\setminus (B_1\cup B_2)$ with $w(v)=1$.
\begin{enumerate}[(a)]
	\item \label{SituationA}	If $v$ has a neighbor in $C$, then this neighbor receives $\frac{1}{3}$ and the neighbor in $A\setminus C$ receives $\frac{2}{3}$. \item \label{SituationB} Otherwise, both neighbors in $A\setminus C$ receive $\frac{1}{2}$.
\end{enumerate}

Now, let $v\in B\setminus(B_1\cup B_2)$ with $w(v)=2$.
\begin{enumerate}[(a),resume]
	\item \label{SituationC} If $v$ has degree $1$ to $A\setminus C$, then $v$ sends $\frac{1}{3}$ along each edge to $C$ and $\frac{4}{3}$ to the neighbor in $A\setminus C$. Note that this neighbor must have a weight of $2$ by \cref{cor:single_vertex_improvement_without_C}.
	\item \label{SituationD}If $v$ has degree $2$ to $A\setminus C$, $v$ sends $1$ along each edge to a vertex in $A\setminus C$ of weight $2$, $\frac{2}{3}$ along each edge to a vertex in $A\setminus C$ of weight $1$, and the remaining amount to the neighbor in $C$.
	\item \label{SituationE}If all three incident edges of $v$ connect to $A\setminus C$, then $v$ sends $\frac{2}{3}$ along each of these edges.
\end{enumerate} 
We denote the set of vertices to which case $\ell$ with $\ell\in\{a,b,c,d,e\}$ applies by $B_\ell$.
\end{definition}
See \cref{FigSecondStep} for an illustration.
\subsection{No Set in $C$ Receives More than $\nicefrac{4}{3}$ Times Its Weight}
\begin{lemma}
	Let $v\in B_d$ and let $u\in N(v,C)$ be the unique neighbor of $v$ in $C$. If $u$ receives more than $\frac{1}{3}$ from $v$, then $w(u)=2$ and $u$ has exactly one incident edge to $B\setminus (B_1\cup B_2)$.\label{LemMoreThanOneThird}
\end{lemma}
\begin{proof}
	Denote the endpoints of the two edges connecting $v$ to $A\setminus C$ by $u_1$ and $u_2$. Assume $u$ receives more than $\frac{1}{3}$ from $v$. Then $w(u_1)=w(u_2)=1$. In particular, $u_1$ and $u_2$ are distinct by \cref{cor:single_vertex_improvement_without_C}. Apply \cref{LemFirstStep} to $U\coloneqq \{u\}$ to obtain $X$ subject to \ref{FirstStepProp1}-\ref{FirstStepProp2}. Then by \ref{FirstStepProp2}, $Y\coloneqq X\dot{\cup}\{v\}$ is a disjoint collection of sets. Moreover, \cref{prop:degree_in_conflict_graph} yields \[|X|\stackrel{\ref{FirstStepProp2}}{=}|N(u,B_1\cup B_2)|\leq |u|\leq 3.\]  Hence, $|Y|\leq 4$. By \ref{FirstStepProp3} and as $u\in C$ receives its own weight in the first step, we get $w(u)=w(X)$. Thus, $w(u_1)+w(u_2)=1+1=2=w(v)$ results in
	\[w(N(Y,A))\stackrel{\ref{FirstStepProp1}}{=}w(u)+w(u_1)+w(u_2)=w(X)+w(v)=w(Y).\] As $Y$ does not constitute a local improvement, $N(Y,A)=\{u_1,u_2,u\}$ contains at least as many vertices of weight $2$ as $Y$. As $w(u_1)=w(u_2)=1$, but $w(v)=2$, this implies that $w(u)=2$ and that all elements of $X$ have a weight of $1$. By \ref{FirstStepProp3}, this implies $|X|=2$, and by \ref{FirstStepProp2}, $u$ intersects sets from $B_1\cup B_2$ in at least two distinct elements in total. In particular, $\{u,v\}$ is the only edge connecting $u$ to $B\setminus (B_1\cup B_2)$ by \cref{prop:degree_in_conflict_graph}. 
\end{proof}
\begin{lemma}\label{lemma:sets_in_C}
	Each set in $C$ receives at most $\frac{4}{3}$ times its own weight during our weight distribution.
\end{lemma}
\begin{proof}
	First, let $u\in C$ with $w(u)=1$. Then $u$ receives $1$ in the first step and has at most one incident edge to $B\setminus (B_1\cup B_2)$. Via this edge, $u$ receives at most $\frac{1}{3}$, which is clear for the cases \ref{SituationA} and \ref{SituationC}, and follows from \cref{LemMoreThanOneThird} for case \ref{SituationD}. Thus, $u$ receives at most $\frac{4}{3}=\frac{4}{3}\cdot w(u)$ in total.
	
	Next, let $u\in C$ with $w(u)=2$. Then $u$ receives $2$ in the first step and $u$ has at most two incident edges to $B\setminus(B_1\cup B_2)$. If $u$ has two incident edges to $B\setminus(B_1\cup B_2)$, then $u$ can receive at most $\frac{1}{3}$ via each of them: This is clear for the cases \ref{SituationA} and \ref{SituationC}, and follows from \cref{LemMoreThanOneThird} for case \ref{SituationD}. Thus, $u$ receives at most $\frac{8}{3}=\frac{4}{3}\cdot w(u)$ in total. If $u$ has one incident edge to $B\setminus (B_1\cup B_2)$, then the maximum amount $u$ can receive via this edge is $\frac{2}{3}$. Again, $u$ receives at most $\frac{8}{3}$ in total.
\end{proof}
\subsection{No Set in $A\setminus C$ Receives More than $\nicefrac{4}{3}$ Times Its Weight}
In order to make sure that no vertex from $A\setminus C$ receives more than $\frac{4}{3}$ times its own weight, we need \cref{LemBoundHighCharges}, which essentially states the following:
\begin{itemize}
	\item If a vertex $u\in A\setminus C$ with $w(u)=2$ receives $\frac{4}{3}$ from a vertex in $B_c$, then it does not receive weight from any further vertex in $B_1\cup B_2\cup B_c\cup B_d$.
	\item A vertex $u\in A\setminus C$ with $w(u)=2$ may, in total, receive at most $2$ units of weight from vertices in $B_1\cup B_2\cup B_d$.
\end{itemize}
\begin{lemma}
	Let $u\in A\setminus C$ with $w(u)=2$. Denote the set of vertices $v\in B_d$ that are connected to $u$ by one/two parallel edges by $D_1$ and $D_2$, respectively.
	
	Then $|N(u,B_1\cup B_2)|+2|N(u,B_c)|+|D_1|+2 |D_2|\leq 2$.\label{LemBoundHighCharges}
\end{lemma}
Our strategy to prove \cref{LemBoundHighCharges} can be summarized as follows: We show that similar to \cref{LemFirstStep}, we can represent the term
$2|N(u,B_c)|+|D_1|+2 |D_2|$ as the weight of a disjoint set collection $Y$ with $N(Y,A \setminus C)\subseteq \{u\}$. $Y$ consists of subsets of sets in $B\setminus (B_1\cup B_2)$. 

We then apply \cref{LemFirstStep} to $U\coloneqq N(Y,C)\cup \{u\}$ to obtain a set collection $X$. We argue that if $|N(u,B_1\cup B_2)|+2|N(u,B_c)|+|D_1|+2 |D_2|> 2=w(u)$, then $X\cup Y$ constitutes a local improvement. In order to arrive at the desired contradiction, we need to initially restrict our attention to a minimal sub-family $\bar{Y}\subseteq N(u,B_c\cup B_d)$ with $|N(u,B_1\cup B_2)|+2|\bar{Y}\cap B_c|+|\bar{Y}\cap D_1|+2 |\bar{Y}\cap D_2| > 2$, which allows us to conclude that $|X\cup Y|\leq 10$.
\begin{proof}[Proof of \cref{LemBoundHighCharges}]
	Assume towards a contradiction that \[|N(u,B_1\cup B_2)|+2|N(u,B_c)|+|D_1|+2 |D_2|\geq 3.\]Note that $|N(u,B_1\cup B_2)|\leq 1$ because $u\not\in C$ and $u$ receives at least one unit of weight per neighbor in $B_1\cup B_2$. Pick an inclusion-wise minimal set $\bar{Y}\subseteq N(u,B_c\cup B_d)$ such that \begin{equation}|N(u,B_1\cup B_2)|+2|\bar{Y}\cap B_c|+|\bar{Y}\cap D_1|+2 |\bar{Y}\cap D_2|\geq 3.\label{eq:size_barY}\end{equation} Then \begin{align}
	|N(u,B_1\cup B_2)|+2|\bar{Y}\cap B_c|+|\bar{Y}\cap D_1|+2 |\bar{Y}\cap D_2|&= 3\text{, or}\label{EqCase1}\\ \bar{Y}\cap D_1=\emptyset\text{ and } |N(u,B_1\cup B_2)|+2|\bar{Y}\cap B_c|+2 |\bar{Y}\cap D_2|&= 4.\label{EqCase2}
	\end{align} 
	We construct a set collection $Y$ as follows: We start with $Y=\emptyset$ and first add all sets contained in $\bar{Y}\cap (B_c\cup D_2)$ to $Y$. Note that for such a set $v$, $N(v,A\setminus C)=\{u\}$ (see \cref{FigSecondStep}). Second, for each $v\in \bar{Y}\cap D_1$, let $v'$ be the set of cardinality $2$ containing the element in which $v$ intersects a set from $C$, and the element in which $v$ intersects $u$. Add $v'$ to $Y$.
	Then $Y$ has the following properties: \begin{align}N(Y,A)&\subseteq C\cup\{u\}\label{eq:neighborhood_Y}\\
	|Y|&=|\bar{Y}\cap B_c| + |\bar{Y}\cap D_1|+|\bar{Y}\cap D_2|\label{eq:size_Y}\\
	w(Y)&=2|\bar{Y}\cap B_c|+|\bar{Y}\cap D_1|+2 |\bar{Y}\cap D_2|\stackrel{\eqref{eq:size_barY}}{\geq} 3-|N(u,B_1\cup B_2)|\label{eq:bound_weight_Y}\\
	|N(Y,C)|&\leq 2|\bar{Y}\cap B_c|+|\bar{Y}\cap D_1|+|\bar{Y}\cap D_2|\label{eq:bound_cardinality_N_Y_C}.\end{align}
	The inequality \eqref{eq:bound_cardinality_N_Y_C} holds since each vertex in $B_c$ has at most $2$ neighbors in $C$, and each vertex in $B_d$ has at most one neighbor in $C$ (see \cref{FigSecondStep}).
	
	Let $U\coloneqq N(Y,C)\cup\{u\}$. Apply Lemma~\ref{LemFirstStep} to obtain $X$ subject to \ref{FirstStepProp1}-\ref{FirstStepProp2}. Then by \ref{FirstStepProp3}, we get \begin{equation}w(X)\geq w(N(Y,C))+|N(u,B_1\cup B_2)|\label{eq:bound_weight_X}\end{equation} because each set in $N(Y,C)$ receives its weight in the first step, and $u$ receives at least one per neighbor in $B_1\cup B_2$. By \ref{FirstStepProp2} and since the sets in $Y$ constitute disjoint subsets of sets in $B\setminus(B_1\cup B_2)$, $X\dot{\cup}Y$ is a family of pairwise disjoint sets. We would like to show that $X\cup Y$ yields a local improvement of size at most $10$.
	By \eqref{eq:bound_weight_X} and \eqref{eq:bound_weight_Y}, we obtain
	\begin{align*}
	w(X\cup Y)&=w(X)+w(Y)\geq 3+w(N(Y,C))\\
	&> w(u)+w(N(Y,C))\geq w(N(X\cup Y,A)),
	\end{align*}
	where $N(X\cup Y,A)\subseteq N(Y,C)\cup \{u\}$ follows from \ref{FirstStepProp1} and \eqref{eq:neighborhood_Y}.
	Thus, it remains to show that $|X\cup Y|\leq 10$. By \ref{FirstStepProp2}, we have
	\begin{align}|X|&=|N(U,B_1\cup B_2)|\leq |N(u,B_1\cup B_2)|+|N(N(Y,C),B_1\cup B_2)|\notag\\
	&\leq|N(u,B_1\cup B_2)|+2 |N(Y,C)|.\label{eq:bound_card_X}\end{align}
	For the last inequality, we used \cref{prop:degree_in_conflict_graph}, which tells us that each set \mbox{$z\in N(Y,C)$} has degree at most $3$ in $G$. In addition, $z$ must intersect at least one set from $Y$, and thus, from $\bar{Y}$. In particular, $z$ has at least one incident edge to $B\setminus (B_1\cup B_2)\supseteq \bar{Y}$, and, thus, at most two incident edges to $B_1\cup B_2$.
	Hence, we obtain 
	\begin{align*}|Y|+|X|&\stackrel{\eqref{eq:bound_card_X}}{\leq }|Y|+|N(u,B_1\cup B_2)|+2 |N(Y,C)|\\
	&\overset{\eqref{eq:size_Y}}{\underset{\eqref{eq:bound_cardinality_N_Y_C}}{\leq}} \underbrace{|N(u,B_1\cup B_2)|+ 5|\bar{Y}\cap B_c|+3|\bar{Y}\cap D_1|+3|\bar{Y}\cap D_2|}_{\eqqcolon (*)}.\end{align*}
	 If \eqref{EqCase1} holds, we can bound $(*)$ by $3$ times the right-hand side of \eqref{EqCase1} and deduce an upper bound of $9$.
	In case \eqref{EqCase2} is satisfied, we can bound $(*)$ by $\frac{5}{2}$ times the right-hand side of \eqref{EqCase2} and obtain an upper bound of $10$. Thus, we have found a local improvement of size at most $10$, a contradiction.
	
\end{proof}
\begin{lemma}\label{lemma:sets_not_in_C}
	Each set $u\in A\setminus C$ receives at most $\frac{4}{3}$ times its own weight during our weight distribution.
\end{lemma}
\begin{proof}
	If $w(u)=1$, then $u$ cannot receive any weight in the first step because otherwise, it would receive at least $1$ and be contained in $C$. Moreover, $u$ has at most two incident edges and receives at most $\frac{2}{3}$ via either of them in the second step.
	
	Next, consider the case where $w(u)=2$. If $u$ receives $\frac{4}{3}$ from a vertex in $B_c$, then by \cref{LemBoundHighCharges}, there is no further vertex in $B_1\cup B_2\cup B_c\cup B_d$ from which $u$ receives weight. As $u$ receives at most $\frac{2}{3}$ per edge in all remaining cases, $u$ receives at most $\frac{4}{3}+2\cdot\frac{2}{3}=\frac{8}{3}=\frac{4}{3}\cdot w(u)$.
	Finally, assume that $N(u,B_c)=\emptyset$. In the first step, $u$ can receive at most $1$ in total (otherwise, $u\in C$) and this can only happen if $u$ has a neighbor in $B_1\cup B_2$. The maximum amount $u$ can receive through one edge in the second step is $1$, and this can only happen in situation \ref{SituationD}. By Lemma~\ref{LemBoundHighCharges}, there are at most $2$ edges via which $u$ receives $1$. Moreover, $u$ can receive at most $\frac{2}{3}$ via the remaining edges. Again, we obtain an upper bound of $1+1+\frac{2}{3}=\frac{8}{3}$ on the total weight received.	
\end{proof}
Combining \cref{lemma:sets_in_C} and \cref{lemma:sets_not_in_C} proves \cref{TheoMainHereditary}. Together with \cref{prop:poly_time} and \cref{TheoFernandesLintzmayer}, we obtain \cref{cor:4_3_MLSA}.
\begin{corollary}\label{cor:4_3_MLSA}
	There is a polynomial-time $\frac{4}{3}$-approximation algorithm for the MLSA in dags.
\end{corollary}
\section{Conclusion}
In this paper, we have presented a simple local search-based $\frac{4}{3}$-approximation for the MLSA in dags, improving upon the previous state-of-the-art of $\frac{7}{5}$ due to Fernandes and Lintzmayer~\cite{FernandesLintzmayer}. Our result is based on a reduction to the hereditary $3$-set packing problem given in~\cite{FernandesLintzmayer}. Given that in~\cite{FernandesLintzmayer}, the reduction is performed in a rather complicated ad-hoc fashion requiring several pages of analysis, the connection between the MLSA in dags and the hereditary $3$-set packing problem remains rather opaque. In this work, we have shown via a very simple reduction that the MLSA in dags is, at its core, a hereditary set packing problem. We have further explored the general connection between approximation guarantees for the hereditary set packing problem and its restriction to instances with bounded set sizes. More precisely, we have seen that an $\alpha$-approximation algorithm for the hereditary $k$-set packing problem implies a $\max\{\alpha,\frac{k+1}{k}\}$-approximation for the hereditary set packing problem. The relation between approximation guarantees for the hereditary $3$-set packing problem and the MLSA in dags obtained by Fernandes and Lintzmayer~\cite{FernandesLintzmayer} corresponds to the special case $k=3$.

Finally, we have established a lower bound of $2-\frac{2}{k}$ on the approximation guarantee achieved by a local search algorithm for the hereditary $k$-set packing problem that only considers local improvements of constant size.

As a result, we can conclude that the approximation guarantee of $\frac{4}{3}$ is best possible for the type of algorithm we consider.

Whether a better guarantee than $\frac{4}{3}$ can be, for example, obtained via a reduction to the hereditary $k$-set packing problem with $k\geq 4$ and an algorithm that considers local improvements of super-constant size remains a question for future research. Note that the state-of-the-art approximation algorithms for the unweighted $k$-set packing problem crucially rely on also considering well-structured local improvements of logarithmic size~\cite{Cygan,FurerYu}.

Finally, it would be interesting to see whether there are other problems that can, in a natural way, be interpreted as a special type of set packing problem that allows for improved approximation guarantees.
\subsubsection*{Acknowledgements.} 
Meike Neuwohner was supported by the Engineering and Physical Sciences Research Council, part of UK Research and Innovation, grant ref.\ EP/X030989/1.
\subsubsection*{Data availability statement.}
No data are associated with this article. Data sharing is not applicable to this article.
\FloatBarrier
\bibliographystyle{plainurl}
\bibliography{arborescences_many_leaves}

\begin{thebibliography}{10}

\bibitem{SpanningDirectedTreesWithManyLeaves}
Noga Alon, Fedor~V. Fomin, Gregory Gutin, Michael Krivelevich, and Saket
  Saurabh.
\newblock Spanning directed trees with many leaves.
\newblock {\em SIAM Journal on Discrete Mathematics}, 23(1):466--476, 2009.
\newblock \href {https://doi.org/10.1137/070710494}
  {\path{doi:10.1137/070710494}}.

\bibitem{KernelsForProblemsWithNoKernel}
Daniel Binkele-Raible, Henning Fernau, Fedor~V. Fomin, Daniel Lokshtanov, Saket
  Saurabh, and Yngve Villanger.
\newblock Kernel(s) for problems with no kernel: On out-trees with many leaves.
\newblock {\em ACM Trans. Algorithms}, 8(4), 2012.
\newblock \href {https://doi.org/10.1145/2344422.2344428}
  {\path{doi:10.1145/2344422.2344428}}.

\bibitem{BONSMA201214}
Paul Bonsma.
\newblock Max-leaves spanning tree is {APX}-hard for cubic graphs.
\newblock {\em Journal of Discrete Algorithms}, 12:14--23, 2012.
\newblock \href {https://doi.org/10.1016/j.jda.2011.06.005}
  {\path{doi:10.1016/j.jda.2011.06.005}}.

\bibitem{Cygan}
Marek Cygan.
\newblock {I}mproved {A}pproximation for 3-{D}imensional {M}atching via
  {B}ounded {P}athwidth {L}ocal {S}earch.
\newblock In {\em 54th Annual {IEEE} Symposium on Foundations of Computer
  Science, {FOCS} 2013, 26-29 October, 2013, Berkeley, CA, {USA}}, pages
  509--518. {IEEE} Computer Society, 2013.
\newblock \href {https://doi.org/10.1109/FOCS.2013.61}
  {\path{doi:10.1109/FOCS.2013.61}}.

\bibitem{DaligaultThomasse}
Jean Daligault and St{\'e}phan Thomass{\'e}.
\newblock On finding directed trees with many leaves.
\newblock In Jianer Chen and Fedor~V. Fomin, editors, {\em Parameterized and
  Exact Computation}, pages 86--97. Springer Berlin Heidelberg, 2009.
\newblock \href {https://doi.org/10.1007/978-3-642-11269-0_7}
  {\path{doi:10.1007/978-3-642-11269-0_7}}.

\bibitem{Drescher2010AnAA}
Matthew Drescher and Adrian Vetta.
\newblock An approximation algorithm for the maximum leaf spanning arborescence
  problem.
\newblock {\em ACM Trans. Algorithms}, 6(3), 2010.
\newblock \href {https://doi.org/10.1145/1798596.1798599}
  {\path{doi:10.1145/1798596.1798599}}.

\bibitem{edmonds1965maximum}
Jack Edmonds.
\newblock Maximum matching and a polyhedron with 0,1-vertices.
\newblock {\em Journal of Research of the National Bureau of Standards Section
  B Mathematics and Mathematical Physics}, 69B:125--130, 1965.
\newblock \href {https://doi.org/10.6028/jres.069b.013}
  {\path{doi:10.6028/jres.069b.013}}.

\bibitem{ErdosSachs}
Paul Erd{\H{o}}s and Horst Sachs.
\newblock Regul{\"a}re {G}raphen gegebener {T}aillenweite mit minimaler
  {K}notenzahl.
\newblock {\em Wiss. Z. Martin-Luther-Univ. Halle-Wittenberg Math.-Natur.
  Reihe}, 12(3):251--257, 1963.

\bibitem{FERNANDES2022217}
Cristina~G. Fernandes and Carla~N. Lintzmayer.
\newblock Leafy spanning arborescences in dags.
\newblock {\em Discrete Applied Mathematics}, 323:217--227, 2022.
\newblock \href {https://doi.org/10.1016/j.dam.2021.06.018}
  {\path{doi:10.1016/j.dam.2021.06.018}}.

\bibitem{FernandesLintzmayer}
Cristina~G. Fernandes and Carla~N. Lintzmayer.
\newblock How heavy independent sets help to find arborescences with many
  leaves in dags.
\newblock {\em Journal of Computer and System Sciences}, 135:158--174, 2023.
\newblock \href {https://doi.org/https://doi.org/10.1016/j.jcss.2023.02.006}
  {\path{doi:https://doi.org/10.1016/j.jcss.2023.02.006}}.

\bibitem{FurerYu}
Martin F{\"{u}}rer and Huiwen Yu.
\newblock Approximating the $k$-{S}et {P}acking {P}roblem by {L}ocal
  {I}mprovements.
\newblock In {\em International Symposium on Combinatorial Optimization}, pages
  408--420. Springer, 2014.
\newblock \href {https://doi.org/10.1007/978-3-319-09174-7_35}
  {\path{doi:10.1007/978-3-319-09174-7_35}}.

\bibitem{GALBIATI199445}
G.~Galbiati, F.~Maffioli, and A.~Morzenti.
\newblock A short note on the approximability of the maximum leaves spanning
  tree problem.
\newblock {\em Information Processing Letters}, 52(1):45--49, 1994.
\newblock \href {https://doi.org/10.1016/0020-0190(94)90139-2}
  {\path{doi:10.1016/0020-0190(94)90139-2}}.

\bibitem{GareyJohnson}
Michael~R. Garey and David~S. Johnson.
\newblock {\em Computers and Intractability; A Guide to the Theory of
  NP-Completeness}.
\newblock W. H. Freeman \& Co., USA, 1990.

\bibitem{Guha1998}
S.~Guha and S.~Khuller.
\newblock Approximation algorithms for connected dominating sets.
\newblock {\em Algorithmica}, 20(4):374--387, 1998.
\newblock \href {https://doi.org/10.1007/PL00009201}
  {\path{doi:10.1007/PL00009201}}.

\bibitem{karp1972reducibility}
Richard~M. Karp.
\newblock Reducibility among combinatorial problems.
\newblock In Raymond~E. Miller, James~W. Thatcher, and Jean~D. Bohlinger,
  editors, {\em Complexity of Computer Computations: Proceedings of a symposium
  on the Complexity of Computer Computations}. Plenum Press, 1972.
\newblock \href {https://doi.org/10.1007/978-1-4684-2001-2_9}
  {\path{doi:10.1007/978-1-4684-2001-2_9}}.

\bibitem{OnSyntacticVersusComputationalViewsOfApproximability}
Sanjeev Khanna, Rajeev Motwani, Madhu Sudan, and Umesh Vazirani.
\newblock On syntactic versus computational views of approximability.
\newblock {\em SIAM Journal on Computing}, 28(1):164--191, 1998.
\newblock \href {https://doi.org/10.1137/S0097539795286612}
  {\path{doi:10.1137/S0097539795286612}}.

\bibitem{Neuwohner23}
Meike Neuwohner.
\newblock Passing the limits of pure local search for weighted k-set packing.
\newblock In {\em Proceedings of the 2023 Annual ACM-SIAM Symposium on Discrete
  Algorithms (SODA)}, pages 1090--1137. Society for Industrial and Applied
  Mathematics, 2023.
\newblock \href {https://doi.org/10.1137/1.9781611977554.ch41}
  {\path{doi:10.1137/1.9781611977554.ch41}}.

\bibitem{RUAN2004325}
Lu~Ruan, Hongwei Du, Xiaohua Jia, Weili Wu, Yingshu Li, and Ker-I Ko.
\newblock A greedy approximation for minimum connected dominating sets.
\newblock {\em Theoretical Computer Science}, 329(1-3):325--330, 2004.
\newblock \href {https://doi.org/10.1016/j.tcs.2004.08.013}
  {\path{doi:10.1016/j.tcs.2004.08.013}}.

\bibitem{SchwartgesSpoerhaseWolff}
Nadine Schwartges, Joachim Spoerhase, and Alexander Wolff.
\newblock Approximation algorithms for the maximum leaf spanning tree problem
  on acyclic digraphs.
\newblock In Roberto Solis-Oba and Giuseppe Persiano, editors, {\em
  Approximation and Online Algorithms}, pages 77--88. Springer Berlin
  Heidelberg, 2012.
\newblock \href {https://doi.org/10.1007/978-3-642-29116-6_7}
  {\path{doi:10.1007/978-3-642-29116-6_7}}.

\bibitem{SolisOba2015A2A}
Roberto Solis-Oba, Paul~S. Bonsma, and Stefanie Lowski.
\newblock A 2-approximation algorithm for finding a spanning tree with maximum
  number of leaves.
\newblock {\em Algorithmica}, 77:374--388, 2015.
\newblock \href {https://doi.org/10.1007/s00453-015-0080-0}
  {\path{doi:10.1007/s00453-015-0080-0}}.

\bibitem{ThieryWard}
Theophile Thiery and Justin Ward.
\newblock An improved approximation for maximum weighted k-set packing.
\newblock In {\em Proceedings of the 2023 Annual ACM-SIAM Symposium on Discrete
  Algorithms (SODA)}, pages 1138--1162. Society for Industrial and Applied
  Mathematics, 2023.
\newblock \href {https://doi.org/10.1137/1.9781611977554.ch42}
  {\path{doi:10.1137/1.9781611977554.ch42}}.

\bibitem{Trevisan}
Luca Trevisan.
\newblock Non-approximability results for optimization problems on bounded
  degree instances.
\newblock In {\em Proceedings of the Thirty-Third Annual ACM Symposium on
  Theory of Computing}, STOC '01, page 453–461, New York, NY, USA, 2001.
  Association for Computing Machinery.
\newblock \href {https://doi.org/10.1145/380752.380839}
  {\path{doi:10.1145/380752.380839}}.

\end{thebibliography}
\end{document}